\documentclass[aps,superscriptaddress,amsmath,amssymb,floatfix,twocolumn,showpacs,amsfonts,longbibliography]{revtex4-1}
\usepackage{times}
\usepackage[varg]{txfonts}
\usepackage{textcomp}
\usepackage{graphicx}
\usepackage{subfigure}
\usepackage{color}
\usepackage[colorlinks=true,citecolor=blue,urlcolor=blue,linkcolor=blue,hyperindex]{hyperref}
\usepackage{braket}
\usepackage{overpic}

\def\pp{\mathbf{p}}
\def\kk{\mathbf{k}}
\def\qq{\mathbf{q}}
\def\rr{\mathbf{r}}
\def\bs{\mathbf{S}}

\def\bq{\mathbf{Q}}
\def\bdelta{\boldsymbol{\delta}}
\def\non{\nonumber}
\def\half{\frac{1}{2}}

\allowdisplaybreaks

\begin{document}

\title{Signatures of indirect K$-$edge resonant inelastic x-ray scattering on magnetic excitations in triangular lattice antiferromagnet}
\author{Cheng Luo}
\affiliation{State Key Laboratory of Optoelectronic Materials and Technologies, School of Physics and Engineering, Sun Yat-Sen University, Guangzhou 510275, China}

\author{Trinanjan Datta}
\email[Corresponding author:]{tdatta@gru.edu}
\affiliation{Department of Chemistry and Physics, Georgia Regents University, 1120 15th Street, Augusta, Georgia 30912, USA}

\author{Zengye Huang}
\affiliation{State Key Laboratory of Optoelectronic Materials and Technologies, School of Physics and Engineering, Sun Yat-Sen University, Guangzhou 510275, China}

\author{Dao-Xin Yao}
\email[Corresponding author:]{yaodaox@mail.sysu.edu.cn}
\affiliation{State Key Laboratory of Optoelectronic Materials and Technologies, School of Physics and Engineering, Sun Yat-Sen University, Guangzhou 510275, China}

\date{\today}

\begin{abstract}
We compute the  K$-$edge \textit{indirect} resonant inelastic x-ray scattering (RIXS) spectrum of a triangular lattice antiferromagnet
in its ordered coplanar 3$-$ sublattice $120^\circ$ magnetic state.
Conventional K$-$edge RIXS spectra prohibits the presence of odd spin flip terms. However noncollinearity of the spin arrangement in a triangular lattice causes
the transverse and longitudinal spin components to be coupled giving rise to intrinsic odd spin flip trimagnon excitations.
By considering the first order self$-$energy corrections to the spin wave spectrum, magnon decay rate, bimagnon interactions within the ladder approximation Bethe-Salpeter scheme, and the effect of three$-$magnon contributions up to $1/S-$ order we find that the RIXS spectra is non$-$trivially affected. For a purely isotropic triangular lattice model, the peak splitting mechanism and the appearance of a multipeak RIXS structure is primarily dictated by the damping of magnon modes. At a scattering wavevector corresponding to the zone center $\Gamma$ point and at the roton point $\qq=M$ where the magnon decay rate is zero a stable single peak forms. However, the microscopic origins of these peaks are different. At the $\Gamma$ point, the contribution is purely trimagnon at the $1/S$ level and occurs approximately at the trimagnon energy of $6JS$. This provides experimentalists with a means to detect purely trimagnon excitations, even at the K$-$edge.
The roton peak occurs at a lower energy of $4JS$. The K$-$edge single peak RIXS spectra at the roton momentum can be utilized
as an experimental signature to detect the presence of roton excitations.
A unique feature of the triangular lattice K$-$edge RIXS spectra is the nonvanishing RIXS intensity at both the zone center $\Gamma$ point and the antiferromagnetic wavevector $K$ point. This result is in sharp contrast to the vanishing K$-$edge RIXS intensity of the collinear ordered magnetic phases on the square lattice. We find that including $XXZ$ anisotropy leads to additional peak splitting, including at the roton scattering wavevector where the single peak destabilizes towards a two$-$peak structure. The observed splitting is consistent with our earlier theoretical
prediction of the effects of spatial anisotropy on the RIXS spectra of a frustrated quantum magnet [Luo, Datta, and Yao,
Phys. Rev. B 89, 165103 (2014)]. In summary, the features of an indirect K$-$edge RIXS spectra of a triangular
lattice quantum magnet can be interpreted as a combination of magnon decay and spin anisotropy effects.

\begin{description}
\item[PACS number(s)] 78.70.Ck, 75.25.-J, 75.10.Jm
\end{description}
\end{abstract}

\maketitle
\section{Introduction}
Contrary to the historical prediction of the spin$-$ $1\over 2$ triangular lattice antiferromagnet (TLAF) as a canonical example of
a spin liquid state~\cite{Anderson1973153}, extensive $-$ theoretical~\cite{PhysRevB.40.2727,JPSJ.61.983,JPCM.6.8891},
numerical~\cite{PhysRevLett.82.3899,PhysRevB.74.224420,PhysRevLett.99.127004,LiPhysRevB.91.014426,HusePhysRevLett.60.2531,
BernuPhysRevB.50.10048,SinghPhysRevLett.68.1766}, and experimental~\cite{PhysRevLett.108.057205,PhysRevB.91.024410,SusukiPhysRevLett.110.267201,OnoPhysRevB.67.104431,KadowakiJPSJ,
IshiiEPL,PetitPhysRevB.81.104411,TothPhysRevB.84.054452, TothPhysRevLett.109.127203} $-$ studies
on the nearest$-$neighbor Heisenberg model has established the ground state configuration as a 120$^{\text{o}}$ long$-$range coplanar 3$-$ sublattice arrangement.
The predicted ordering pattern persists for all values of spin $S$, including the $S = \half$ state where quantum fluctuations lead to a 60\% suppression
of the magnetic order parameter from its classical N\'{e}el value~\cite{PhysRevLett.82.3899,PhysRevB.74.224420,PhysRevLett.99.127004,LiPhysRevB.91.014426}.
At present there exists a plethora of real TLAF materials, with both isotropic and anisotropic interactions, which provide a motivation to study triangular lattice frustrated magnets ~\cite{PhysRevLett.108.057205,PhysRevLett.86.1335,OnoPhysRevB.67.104431,KadowakiJPSJ,IshiiEPL,PetitPhysRevB.81.104411,
TothPhysRevB.84.054452,TothPhysRevLett.109.127203}.
Further impetus to investigate and delineate the physical properties of the TLAF stems from the flurry of recent theoretical and numerical investigation to clarify the ground and excited state properties of both isotropic and anisotropic triangular lattice systems \cite{ChenPhysRevB.87.165123,SchmidtPhysRevB.89.184402,HaukeNJP,nphys749,fishmanPhysRevB.79.184413,
okamotoPhysRevB.81.020402,PhysRevB.91.134423, SuzukiPhysRevB.90.184414, WhitePhysRevB.84.245130,HaukePhysRevB.87.014415}.

Traditionally, information on the magnetic ground state and single$-$magnon excitations is inferred from inelastic neutron scattering (INS)
experiments ~\cite{PhysRevLett.86.5377,PhysRevLett.87.037202}. However, with enhancements in instrumental resolution of the next generation synchrotron radiation sources
resonant inelastic X$-$ray scattering (RIXS) spectroscopy offers the condensed matter and materials science community an alternate option to experimentally probe
magntic excitations in correlated magnets~\cite{RevModPhys.83.705}. As a \emph{photon$-$in} \emph{photon$-$out} spectroscopic technique,
RIXS can offer direct information on both single$-$magnon and multimagnon excitations.
Present efforts to understand the K$-$ edge indirect RIXS spectra are primarily directed towards the study of square lattice compounds in the N\'{e}el
antiferromagnetic and collinear antiferromagnetic phases~\cite{PhysRevLett.100.097001,PhysRevB.81.085124,EPL.80.47003,PhysRevB.75.214414,PhysRevB.77.134428,PhysRevB.89.165103,NJP.14.113038}.
In a recent publication, Ref.~[\onlinecite{PhysRevB.89.165103}], the authors of this paper have shown that in the case of an anisotropic square lattice with strong frustrating further neighbor interactions the RIXS spectrum can split into a robust two$-$peak structure, over a wide range of transferred momenta, in both magnetically ordered phases.
It was also predicted that the unfrustrated model contains a single$-$peak structure.

In RIXS spectroscopy single and three spin flip processes are allowed at the L$-$ and M$-$ edges due to the presence of spin$-$orbit coupling~\cite{PhysRevLett.103.117003,PhysRevLett.105.167404,arXiv:1002.3773}.
But, in a square lattice, excitations of odd magnons are prohibited at the K$-$edge~\cite{RevModPhys.83.705} and the
spectra originates purely from the bimagnon contribution.
In the absence of an external magnetic field the spin ordering in a square lattice system is collinear and
the magnon excitations are long lived without any damping. In contrast, in the TLAF the noncollinear ground state contains inherent three$-$magnon
excitations (odd spin flip terms). The coupling of the longitudinal and transverse spin excitations
gives rise to a finite quasi$-$particle lifetime (see Fig.~\ref{fig:sw}) which introduces an intrinsic damping of the magnon modes ~\cite{RevModPhys.85.219,PhysRevLett.97.207202}.
Hence, the presence of the trimagnon interaction, even at the K$-$edge, motivates several unanswered questions within the context of quantum
magnetism and RIXS spectroscopy. How does the presence of an intrinsic damping affect the indirect K$-$edge RIXS spectra?
What role does the interplay between geometrical frustration and spin anisotropy have on the RIXS spectra?
In this article, we predict the effects of bimagnon and trimagnon processes in indirect RIXS spectroscopy of a geometrically frustrated TLAF,
a topic which is unexplored both theoretically and experimentally.

The microscopic mechanism underlying magnetic excitations in the indirect RIXS process involves a local modification of the superexchange interaction
mediated via the core hole~\cite{EPL.73.121,PhysRevB.75.115118}. The resulting RIXS spectra is expressed as a momentum$-$dependent four-spin correlation function
which can probe bimagnon excitations across the entire Brillouin zone (BZ)~\cite{EPL.80.47003,PhysRevB.75.214414,PhysRevB.77.134428}.
Hence, RIXS is complementary to optical Raman scattering which is restricted to zero momentum~\cite{RevModPhys.79.175,JPCM.19.145243,PhysRevB.77.174412,PhysRevB.87.174423}.
From a theoretical perspective elucidating the nature of the bimagnon dispersion affected both by two$-$magnon ladder scattering processes and three$-$magnon interactions
is challenged by the appearance of several non$-$trivial dynamical properties in the spin$-$wave excitation spectrum ~\cite{PhysRevLett.96.057201,PhysRevLett.97.207202,PhysRevB.74.180403,PhysRevB.79.144416}.
Namely $-$ (i) strong renormalization of magnon energies with respect to the linear spin$-$wave theory result,
(ii) finite lifetime due to spontaneous magnon decays at zero temperature,
and (iii) appearance of rotonlike minima at the edge center of the BZ ($M$ point, see Fig.~\ref{fig:sw}(a)).

The objective of this paper is to elucidate the role of magnon$-$magnon interaction, spontaneous magnon decays, the effect of the rotonlike minima, and spin anisotropy on the indirect RIXS spectra of a TLAF.  For this purpose, we consider an easy$-$plane $XXZ$ triangular lattice model. In the isotropic limit, the $XXZ$ model can provide an accurate description of the Ba$_3$CoSb$_2$O$_9$ system~\cite{PhysRevLett.108.057205,PhysRevLett.109.267206}. In addition, it provides a starting point for the discussion of RIXS effects in anisotropic TLAF \cite{OnoPhysRevB.67.104431,KadowakiJPSJ,IshiiEPL,PetitPhysRevB.81.104411,TothPhysRevB.84.054452,TothPhysRevLett.109.127203}.
We compute the RIXS intensity utilizing the $1/S$ spin$-$wave expansion technique within the Bethe$-$Sapleter scheme where interaction effects arising from
both the quartic terms via the ladder scattering process and the contributions of the cubic anharmonic terms up to $1/S$ order are included.

The main results of our article can be summarized as follows. \textit{First}, in the case of an isotropic nearest neighbor TLAF
we find that the spontaneous magnon decay and kinematic constraints of the phase space inherent to the model is the primary
cause for creating a multipeak (more than two$-$peak) structure in the RIXS spectra. \textit{Second}, contrary to the K$-$edge RIXS intensity of the square lattice case, in the TLAF the RIXS intensity does not vanish at the $\Gamma$ point and at the $K$ point. At the $\Gamma$ point, the bimagnon intensity is zero and the single peak spectra results purely from the trimagnon contribution, approximately at energy scale of 6JS corresponding to the three magnon energy. This provides experimentalists with a means to detect purely trimagnon RIXS spectra at the K$-$edge. At the antiferromagnetic wave vector $K$ point the RIXS intensity is dominated by the bimagnon excitations. \textit{Third}, an important conclusion of our work
is the proposal of utilizing RIXS as a probe to detect the presence of the roton mode. We show that at a scattering wave vector equal to the roton momentum $\qq=M$ the RIXS spectra has a single peak structure. Barring the $\Gamma$ point peak which occurs at a higher energy, at all other special high symmetry points of the magnetic BZ the RIXS spectra splits into a multipeak structure. The appearance of the single peak structure can serve as an experimental signature to detect the appearance of a roton mode in a TLAF. \textit{Fourth}, including the $XXZ$ anisotropy leads to further peak splitting including at the roton scattering wavevector point. \textit{Fifth}, we find that the conceptual signature of slow moving bimagnons as an indicator of RIXS peak splitting (instability), as proposed in our earlier work on the two$-$peak splitting theory within the context of the anisotropic square lattice Heisenberg model~\cite{PhysRevB.89.165103},
still holds (see Fig.~\ref{fig:velgraph}).

This article is organized as follows. In Sec.~\ref{Sec:Model}, we introduce the $XXZ$ Hamiltonian, present the expression for the effective Hamiltonian within an interacting spin$-$wave formalism, and compute the intensity maps for the renormalized energy and magnon decay rate, up to $1/S$ corrections. In Sec.~\ref{Sec:RIXS}, we state the definition and the expression of the TLAF RIXS scattering operator containing both the bimagnon and trimagnon contributions.
In Sec.~\ref{Sec:Results}, we display our results, state the formalism and numerical approach for computing RIXS intensity, and discuss the implications of our result within the context of a TLAF (geometric frustration). First, in Sec.~\ref{Sec:Bare}, we present the results for the noninteracting bimagnon and trimagnon RIXS intensity and spectral weight. In Sec.~\ref{Sec:Vertex}, we outline our formalism and calculate the interacting bimagnon intensity. In Sec.~\ref{Sec:Total}, we calculate the full RIXS spectrum. In Sec.~\ref{Sec:Conclu}, we present our concluding remarks and discuss the appearance of slow moving bimagnons as a signature of peak splitting. Finally, to preserve clarity in the main body of the text, we state the details of the spin$-$wave theory derivation of the effective Hamiltonian in Appendix~\ref{App:SWT}
and display results to validate our numerical approach in Appendix~\ref{Exact solution}.
\section{TLAF MODEL AND MAGNON DECAY}\label{Sec:Model}
\begin{figure}
\centering\includegraphics[scale=0.5]{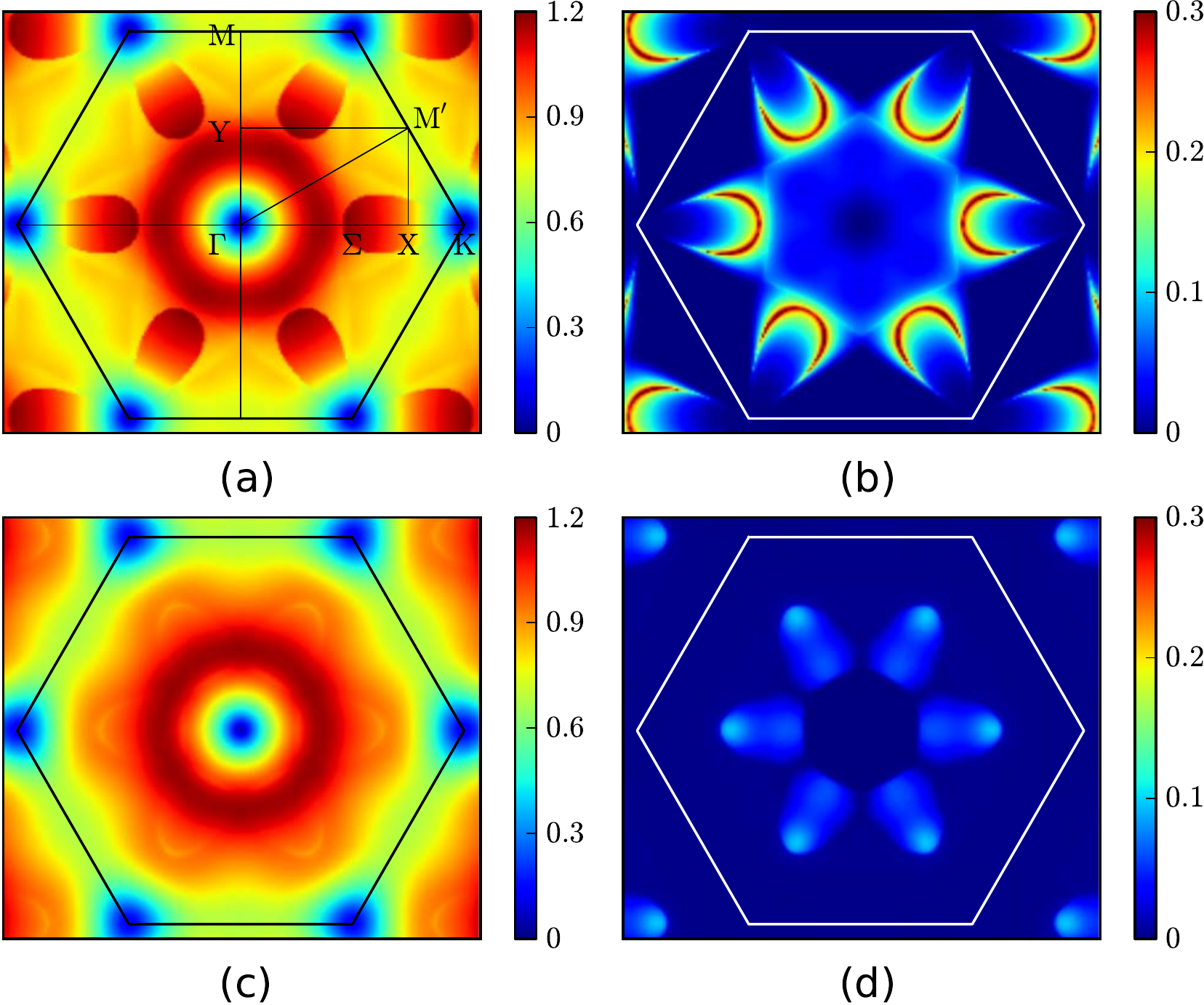}
\caption{(Color online)
Intensity maps of the $1/S$ spin wave spectrum for the $S=1/2$ triangular-lattice antiferromagnet
with easy$-$plane anisotropy $\alpha=1$ (upper row) and $\alpha=0.95$ (lower row).
(a) and (c), renormalized magnon energy $\bar{\varepsilon}_\kk$. (b) and (d), magnon decay rate $\Gamma_\kk$.
$\Gamma=(0,0)$, $\Sigma=(2\pi/3,0)$, $X=(\pi,0)$, $K=(4\pi/3,0)$, $M=(0,2\pi/\sqrt{3})$, and $Y=(0,\pi/\sqrt{3})$ points in (a) are highlighted. Note, the overall damping of spin waves is strongly reduced and the decay region shrinks with increasing anisotropy. The realistic magnon decays disappear at $\alpha\approx0.92 $~\cite{PhysRevB.79.144416}.
}
\label{fig:sw}
\end{figure}
Inelastic neutron scattering data of a TLAF reveals well$-$defined sharp modes in the low$-$energy excitation spectrum accompanied with a broad continuum at intermediate and high energies~\cite{PhysRevB.68.134424,PhysRevLett.109.267206,PhysRevLett.111.257202}.
A number of competing theoretical proposals, ranging from a proximate spin$-$liquid phase~\cite{nphys749,PhysRevB.91.134423} to enhanced magnon$-$magnon interactions~\cite{PhysRevB.72.134429,PhysRevB.73.184403,PhysRevB.88.094407}
have been proposed to explain the nature of the spin wave excitation spectrum.
Our starting point is the spin $S$, nearest$-$neighbor $XXZ$ antiferromagnetic  model on the triangular lattice.
The spin$-$wave theory Hamiltonian in the local $(x-z)$ rotating frame associated
with the ordering wave vector $\bq=(4\pi/3,0)$, $K-$ point in BZ, takes the following form~\cite{PhysRevB.79.144416}:
\begin{eqnarray}\label{localham}
 \mathcal{H}=&&J\sum_{\langle ij\rangle}[\alpha S_i^yS_j^y+\cos(\theta_i-\theta_j)(S_i^zS_j^z+S_i^xS_j^x)\non\\
 &&+\sin(\theta_i-\theta_j)(S_i^zS_j^x-S_i^xS_j^z)].
\end{eqnarray}
where $\theta_i=\bq\cdot\rr_i$ and we have also introduced an easy$-$plane anisotropy parameter $\alpha\in[0,1]$.
In Appendix \ref{App:SWT} we outline the derivation of the effective interacting spin$-$wave Hamiltonian $\mathcal{H}_{\mathrm{eff}}$
in the first order $1/S$ expansion with respect to linear spin$-$wave theory. The resulting expression is
\begin{eqnarray}\label{effectiveham}
 \mathcal{H}_{\mathrm{eff}}=&&\sum_\kk(\varepsilon_\kk+\delta\varepsilon_\kk)b_\kk^\dag b_\kk
 +\frac{1}{2!}\sum_{\{\kk_i\}}V_{a}(b_{1}^\dag b_{2}^\dag b_{3}+\mathrm{H.c.})\non\\
 &&+\frac{1}{3!}\sum_{\{\kk_i\}}V_{b}(b_{1}^\dag b_{2}^\dag b_{3}^\dag+\mathrm{H.c.})
 +\sum_{\{\kk_i\}}V_{c}b_{1}^\dag b_{2}^\dag b_{3}b_{4},
\end{eqnarray}
where we have adopted the convention that $1=\kk_1$, $2=\kk_2$, etc, and momentum conservation is assumed for various $\kk$-summations.
The bare magnon dispersion given by the linear spin$-$wave theory is expressed as
\begin{eqnarray}\label{bareek}
  \varepsilon_\kk=3JS\sqrt{(1-\gamma_{\kk})(1+2\alpha\gamma_\kk)},
\end{eqnarray}
with $\gamma_\kk=\frac{1}{3}[\cos k_x+2\cos(k_x/2)\cos(\sqrt{3}k_y/2)]$.
The explicit forms for the interacting vertices $\delta\varepsilon$, $V_{a,b}$, and $V_c$
are detailed in Appendix \ref{App:SWT}. At zero temperature the bare magnon propagator is defined as
\begin{eqnarray}\label{bareg}
\mathrm{G}_0^{-1}(\kk,\omega)=\omega-\varepsilon_\kk+i0^+.
\end{eqnarray}
The first order $1/S$ correction to the magnon energy is determined by the Dyson equation
\begin{eqnarray}
  \omega-\varepsilon_\kk-\Sigma(\kk,\omega)=0,
\end{eqnarray}
with the one-loop self-erengy $\Sigma(\kk,\omega)=\Sigma_{a}(\kk,\omega)+\Sigma_{b}(\kk,\omega)+\Sigma_{c}(\kk)$,
where $\Sigma_{c}(\kk)=\delta\varepsilon_\kk$ is a frequency-independent Hartree-Fock correction,
while $\Sigma_{a,b}(\kk,\omega)$ are calculated as~\cite{JPCM.6.8891,PhysRevLett.97.207202,PhysRevB.74.180403,PhysRevB.79.144416}
\begin{eqnarray}\label{Eq.self-erengy1}
 \Sigma_{a}(\kk,\omega)&=&\frac{1}{2}\sum_\pp \frac{|V_{a}(\pp,\kk-\pp;\kk)|^2}{\omega-\varepsilon_\pp-\varepsilon_{\kk-\pp}+i0^+},\\ \label{Eq.self-erengy2}
 \Sigma_{b}(\kk,\omega)&=&-\frac{1}{2}\sum_\pp\frac{|V_{b}(\pp,-\kk-\pp,\kk)|^2}{\omega+\varepsilon_\pp+\varepsilon_{\kk+\pp}-i0^+} .
\end{eqnarray}
The on-shell solution consists of setting $\omega=\varepsilon_\kk$ in the self-energy (\ref{Eq.self-erengy1}) and (\ref{Eq.self-erengy2})
leads to the following expression for the $1/S$ renormalized spectrum
\begin{eqnarray}
  \omega_\kk\equiv\bar{\omega}_\kk-i\Gamma_\kk=\varepsilon_\kk+\Sigma(\kk,\varepsilon_\kk).
\end{eqnarray}

In Fig.~\ref{fig:sw} we display the intensity maps for the renormalization of magnon energy $\bar{\varepsilon}_\kk$ and the magnon decay rate $\Gamma_\kk$
for the $S=1/2$ triangular antiferromagnet for $\alpha=1$ (isotropic) and $\alpha=0.95$ (anisotropic) case.
From Fig.~\ref{fig:sw}(d) we observe that the magnon decay rate decreases drastically in the presence of anisotropy.
This is due to the reduced phase volume where the kinematic constraint $\varepsilon_\kk=\varepsilon_\pp-\varepsilon_{\kk-\pp}$ in the self-energy (\ref{Eq.self-erengy1}) is satisfied. The magnon decay intensity maps Fig.~\ref{fig:sw}(b) and Fig.~\ref{fig:sw}(d) play an important role in understanding the origins of the multipeak RIXS structure shown in Fig.~\ref{fig:rixsgm} and Fig.~\ref{fig:fullspectra}.

\section{Indirect RIXS Correlator}\label{Sec:RIXS}

In Mott insulating systems, multimagnon excitations can be created dynamically by the presence of the core-hole potential in the intermediate state of indirect RIXS process. The effective scattering operator, in the first order, under the assumption of the ultra-short core-hole life-time
(UCL) expansion is given by~\cite{EPL.80.47003,PhysRevB.77.134428}
\begin{equation}\label{rixsop}
  \mathcal{R}_\qq=J\sum_{i,\bdelta}e^{i\qq\cdot \mathbf{r}_i}\bs_i\cdot\bs_{i+\bdelta},
\end{equation}
where $\mathbf{r}_i$ is the position of the ion absorbing the incident photon and $\bdelta$ denotes the neighboring vectors.
After consecutive Holstein-Primakoff and Bogoliubov transformations,
the magnon creation parts of the RIXS scattering operator can be expressed in terms of the bosonic operators as
\begin{equation}
\mathcal{R}_\qq=\sum_{1+2=\qq}M(1,2)b_{1}^\dag b_{2}^\dag+\sum_{1+2+3=\qq}N(1,2,3)b_{1}^\dag b_{2}^\dag b_{3}^\dag,
\end{equation}
with the bimagnon scattering matrix element expression is given by
\begin{eqnarray}\label{2mop}
 M(1,2)=&&\frac{3JS}{2!}\Big\{[1+\gamma_\qq+(\alpha-\frac{1}{2})(\gamma_{1}+\gamma_{2})](u_{1}v_{2}+v_{1}u_{2})\non\\
 &&-(\alpha+\frac{1}{2})(\gamma_{1}+\gamma_{2})(u_{1}u_{2}+v_{1}v_{2})\Big\},
\end{eqnarray}
and the trimagnon scattering matrix element is given by
\begin{eqnarray}\label{3mop}
  N(1,2,3)=&&\frac{3JS}{3!}i\sqrt{\frac{3}{2S}}\big[(\bar{\gamma}_1-\bar{\gamma}_{2+3}-\frac{1}{4}\bar{\gamma}_\qq)(u_1+v_1)\non\\
  &&\times (u_2v_3+v_2u_3)+(\bar{\gamma}_2-\bar{\gamma}_{1+3}-\frac{1}{4}\bar{\gamma}_\qq)(u_2+v_2)\non\\
  &&\times (u_1v_3+v_1u_3)+(\bar{\gamma}_3-\bar{\gamma}_{1+2}-\frac{1}{4}\bar{\gamma}_\qq)(u_3+v_3)\non\\
  &&\times (u_1v_2+v_1u_2)\big],
\end{eqnarray}
where $u$, $v$ and $\bar{\gamma}$ are defined in Appendix \ref{App:SWT}. The  three-boson term in our theory has no analog in the collinear phases of a square lattice quantum magnet. Note, the corrections from magnon interactions for the trimagnon intensity appear at the $1/S^{2}$ order and are neglected in the remainder of this paper.

The frequency- and momentum-dependent magnetic scattering intensity is related to multimagnon response function via the fluctuation-dissipation theorem.
The full $1/S$ corrections to the indirect RIXS susceptibility is the sum of bimagnon and trimagnon contributions given by
\begin{eqnarray}\label{fullrixs}
I(\qq,\omega)&=&-\frac{1}{\pi}\mathrm{Im}[\chi_2(\qq,\omega)+\chi_3(\qq,\omega)],\nonumber\\
&=&-\frac{1}{\pi}\mathrm{Im}[ \chi_{\mathrm{RIXS}}(\qq,\omega)],
\end{eqnarray}
which involve an interacting two$-$magnon susceptibility $\chi_2(\qq,\omega)$ and a non$-$interacting three$-$magnon susceptibility $\chi_3(\qq,\omega)$.
The susceptibilities can be expressed explicitly from the corresponding multi-magnon Green's function defined as
\begin{eqnarray}\label{Eq:corrfun}
\chi_2(\qq,\omega)&=&\sum_{\kk\kk'}M_{\kk}M_{\kk'}\Pi_{\kk\kk'}(\qq,\omega),\\ \label{Eq:corrfun3}
\chi_3(\qq,\omega)&=&\sum_{\kk\pp;\kk'\pp'}N_{\kk,\pp}N_{\kk',\pp'}\Lambda_{\kk\pp;\kk'\pp'}(\qq,\omega),
\end{eqnarray}
where $\Pi$ and $\Lambda$ are denoted as the bimagnon and trimagnon propagators, respectively.
The momentum-dependent two-magnon and three-magnon Green's function in terms of Bogoliubov quasiparticles are defined as
\begin{eqnarray}\label{Eq:2mgreen}
i\Pi_{\kk\kk'}(\qq,t)&=&\langle\mathcal{T}b_{\kk+\qq}(t)b_{-\kk}(t)b_{\kk^\prime+\qq}^\dag b_{-\kk^\prime}^\dag\rangle,\\ \label{Eq:3mgreen}
i\Lambda_{\kk\pp;\kk'\pp'}(\qq,t)&=&\langle\mathcal{T}b_{\kk}(t)b_{\qq-\kk-\pp}(t)b_{\pp}(t)b_{\kk^\prime}^\dag b_{\qq-\kk^\prime-\pp^\prime}^\dag b_{\pp^\prime}^\dag \rangle,
\end{eqnarray}
where $\mathcal{T}$ is the time-ordering operator and $\langle\cdot\rangle$ is the average of the ground state.
In the following sections, using Eq.~(\ref{Eq:2mgreen}) and Eq.~(\ref{Eq:3mgreen}), we will compute the noninteracting and the interacting RIXS spectra.

\section{RESULTS AND DISCUSSION}\label{Sec:Results}

\subsection{Noninteracting bi$-$ and trimagnon spectra}\label{Sec:Bare}
Using Eqs.~(\ref{2mop})$-$(\ref{3mop}) and applying Wick's theorem to Eq.~(\ref{Eq:2mgreen}) and Eq.~(\ref{Eq:3mgreen}),
we obtain the following expressions for the noninteracting bimagnon ($I_2(\qq,\omega)$) and trimagnon ($I_3(\qq,\omega)$) scattering intensity
\begin{eqnarray}\label{Eq:bare2m}
I_2(\qq,\omega)&=&2\sum_{\kk}M_{\kk+\qq,-\kk}^2\delta(\omega-\varepsilon_{\kk+\qq}-\varepsilon_{\kk}),\\ \label{Eq:bare3m}
I_3(\qq,\omega)&=&6\sum_{\kk,\pp}N_{\kk,\qq-\kk-\pp,\pp}^2\delta(\omega-\varepsilon_{\kk}-\varepsilon_{\qq-\kk-\pp}-\varepsilon_{\pp}).
\end{eqnarray}
\begin{figure}
\centering\includegraphics[scale=0.45]{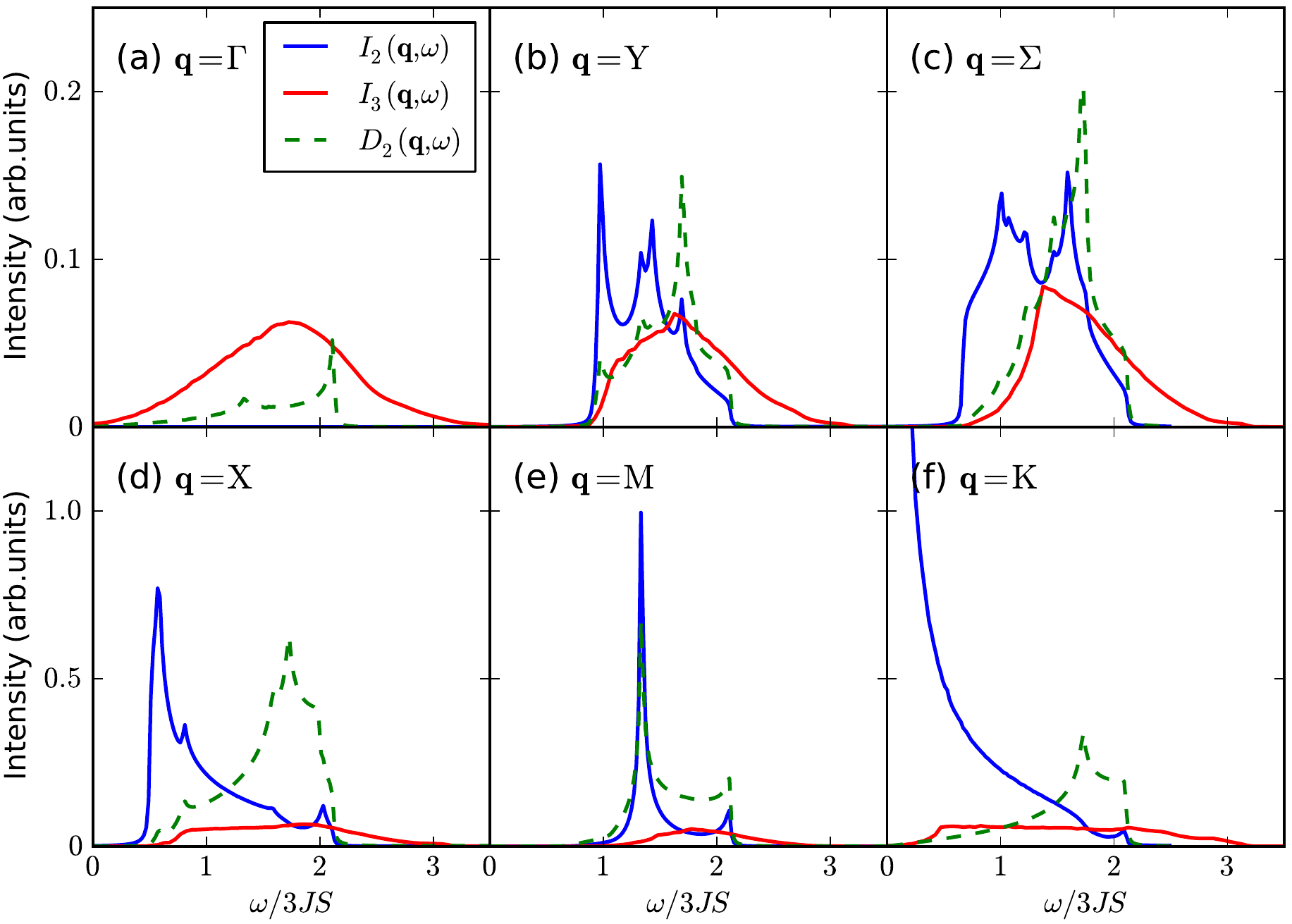}
\caption{(Color online)
Non-interacting bimagnon ($I_2(\qq,\omega)$) and trimagnon ($I_3(\qq,\omega)$) RIXS intensity of the isotropic TLAF
at momentum transfer wavevector $\qq$ corresponding to the special high symmetry points of a triangular lattice magnetic BZ.
The bimagnon DOS $D_2(\qq,\omega)$ is also shown as a dashed line.
The nonzero intensity at the $\Gamma$ and at the $K$ point is a unique RIXS feature of the noncollinear ground state configuration.
}
\label{fig:rixsbare}
\end{figure}
\begin{figure}
\centering\includegraphics[scale=0.5]{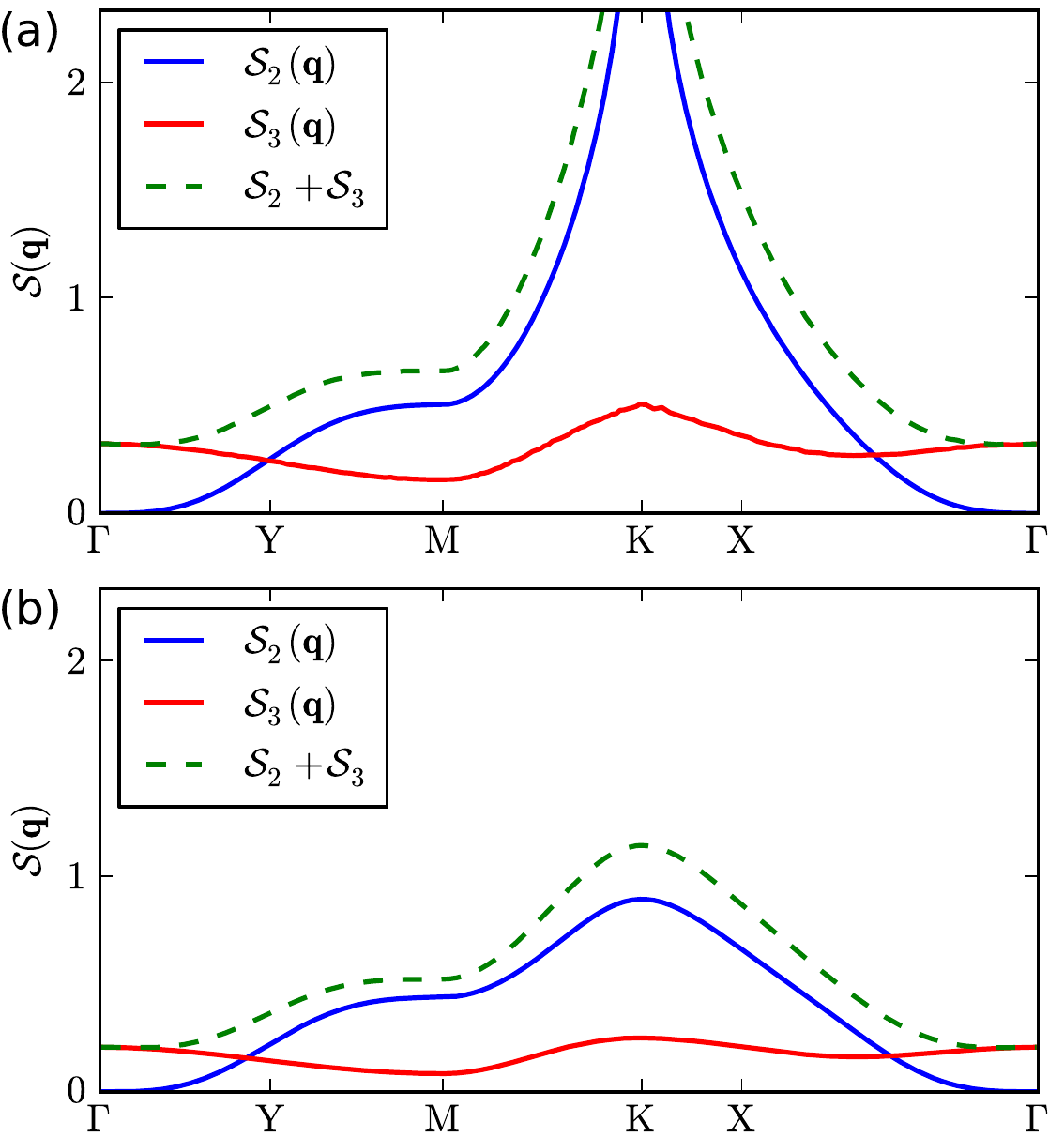}
\caption{(Color online)
The non-interacting bimagnon and trimagnon total spectral weight for the anisotropic TLAF with (a) $\alpha=1$ and (b) $\alpha=0.95$.
Irrespective of the presence of anisotropy, the bimagnon and trimagnon intensity complement each other at the zone center
and at the zone boundary. Introduction of anisotropy removes the singularity due to the opening up of a gap in the spin wave spectrum.
}
\label{fig:rixsweight}
\end{figure}
In Fig.~\ref{fig:rixsbare}, we show the results for the $S=1/2$ isotropic Heisenberg model at various points in the BZ. At the $\Gamma$ point the contribution is purely from the trimagnon excitations, see Fig.~\ref{fig:rixsbare}(a). The bimagnon RIXS intensity displays a nonzero elastic peak at the $K$ point, see Fig.~\ref{fig:rixsbare}(f). The indirect RIXS spectra even at the noninteracting level, in a noncollinear quantum magnet,
exhibits significant differences from the collinear ordered quantum magnets where the intensity vanishes at the BZ
center and at the antiferromagnetic wavevector\cite{EPL.80.47003,PhysRevB.75.214414,PhysRevB.77.134428,PhysRevB.89.165103}.

\begin{figure}
\centering
\includegraphics[scale=0.3]{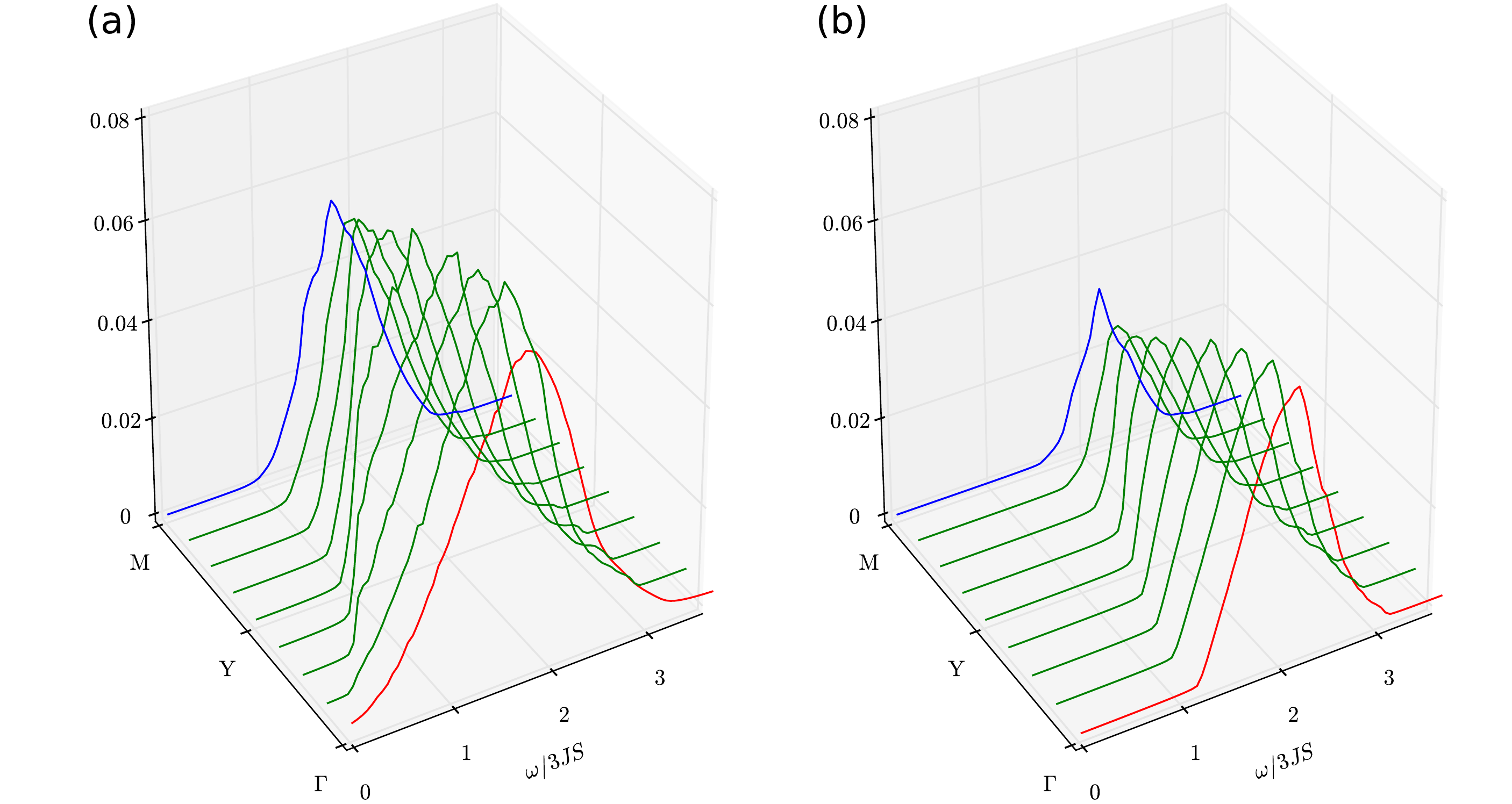}
\caption{(Color online)
Pure trimagnon contribution to the RIXS intensity. (a) no anisotropy  ($\alpha$=1) (b) with anisotropy ($\alpha$=0.95).}
\label{fig:trimagnon}
\end{figure}

The noninteracting bimagnon RIXS intensity in Eq.~\ref{Eq:bare2m} is proportional to the bare two$-$magnon density of states (DOS)
\begin{eqnarray}\label{2mdos}
  D_2(\qq,\omega)=\sum_\kk\delta(\omega-\varepsilon_{\kk+\qq}-\varepsilon_{\kk}),
\end{eqnarray}
A close inspection on the DOS in Fig.~\ref{fig:rixsbare} shows that these Van-Hove singularities which originate from the maximum or saddle points of
the two-magnon continuum $\varepsilon_{\kk+\qq}+\varepsilon_\kk$ partially transfer to the RIXS intensity, see Fig.\ref{fig:rixsbare}(b$-$e),
and the spectrum line shape at $\qq=M$ (Fig.\ref{fig:rixsbare}(e)) resembles the DOS with minimal RIXS matrix element effects.

In Fig.~\ref{fig:rixsweight} we show the variation of the total spectral weight across the BZ for the bimagnon and trimagnon component, respectively. By using the bare intensity Eqs.~(\ref{Eq:bare2m}) and Eqs.~({\ref{Eq:bare3m}) we obtain
\begin{eqnarray}
  \mathcal{S}_2(\qq)&=&\int_0^{\infty}\mathrm{d}\omega I_2(\qq,\omega)=2\sum_\kk M^2(\kk+\qq,-\kk),\\
  \mathcal{S}_3(\qq)&=&\int_0^{\infty}\mathrm{d}\omega I_3(\qq,\omega)=6\sum_\kk N^2(\kk,\qq-\kk-\pp;\pp),
\end{eqnarray}
In general, the trimagnon excitation dominates the indirect RIXS total spectral weight in the vicinity of the BZ center, while the bimagnon spectral weight becomes overwhelmingly large at the boundary of the BZ where the three-magnon intensity is negligible. The most remarkable feature of the isotropic model, see Fig.~\ref{fig:rixsweight}(a), is the elastic peak at the antiferromagnetic wave vector which resembles the longitudinal dynamic structure factor probed by neutron-scattering experiments \cite{PhysRevB.48.3264,PhysRevB.72.224511,PhysRevB.88.094407}. Upon inclusion of anisotropy, $\alpha=0.95$, the elastic peak at $\qq=K$ disappears, see Fig.~\ref{fig:rixsweight}(b), since a gap
is now introduced in the spin$-$wave dispersion (\ref{bareg}) at the ordering wave vector.

In Fig.~\ref{fig:trimagnon} we show the pure trimagnon contribution along the $\Gamma\rightarrow M$ path obtained using the noninteracting three$-$magnon susceptibility $\chi_3(\qq,\omega)$ in Eq.~(\ref{Eq:corrfun3}). We plot the spectra both in the presence and in the absence of anisotropy. We observe that the trimagnon spectra peak occurs at a higher energy approximately around $6JS$ around the $\Gamma$ point, which downshifts before undergoing an upward shift to $6JS$ around the M point. In the presence of anisotropy, see Fig.~\ref{fig:trimagnon}(b), there is an overall upward shift of the energy peak. The observed effect could be an artifact of considering a noninteracting trimagnon spectra. In the next section we consider the interacting bimagnon RIXS intensity up to 1/S order.

\subsection{Bimagnon excitations: $1/S$ corrections}\label{Sec:Vertex}
\begin{figure}
\centering\includegraphics[scale=0.6]{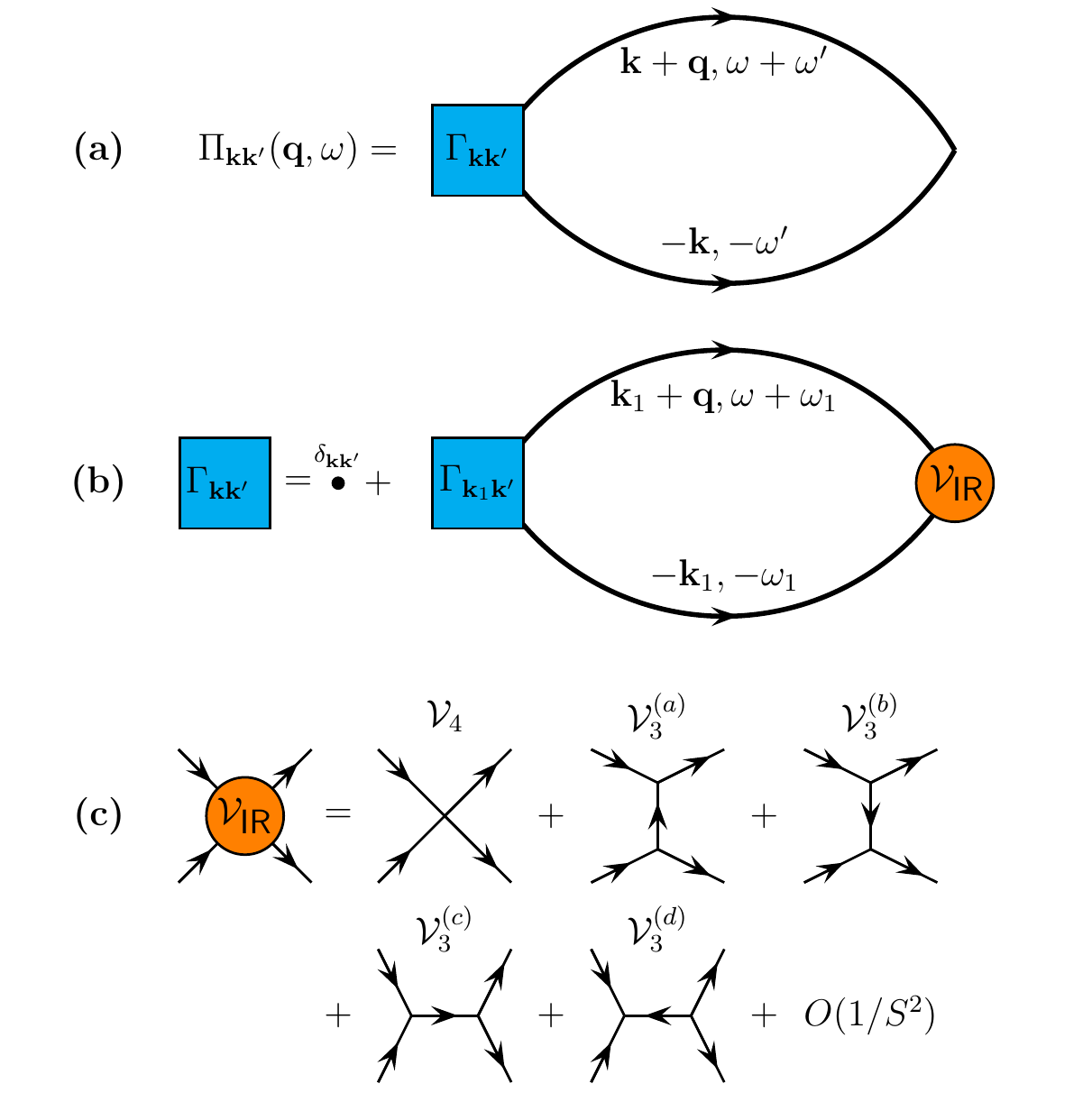}
\caption{(Color online)
Diagrammatic representation for
(a) two-magnon propagator $\Pi_{\kk\kk'}(\qq,\omega)$,
(b) Bethe-Salpeter euqation of the vertex function $\Gamma_{\kk\kk'}(\omega,\omega')$ and
(c) the $1/S$ order irreducible interaction $\mathcal{V}_{\mathrm{IR}}$.
Solid lines with an arrow in (a) and (b) stand for the single-magnon propagators.
}
\label{fig:bethe-salpeter}
\end{figure}
We now proceed with the analysis of $1/S$ correction to the two-magnon Green's function by taking into account both the self-energy correction
to the single magnon propagator $G$ according to the Dyson equation and the vertex insertions to the two-magnon propagator $\Pi$
which satisfies the Bethe-Salpeter (BS) equation~\cite{PhysRevB.4.992,PhysRevB.45.7127}.
The diagrammatic representation of such procedure is depicted in Fig.~\ref{fig:bethe-salpeter}(a) and ~\ref{fig:bethe-salpeter}(b). The total irreducible bimagnon scattering vertices in Fig.~\ref{fig:bethe-salpeter}(c) fall into two categories,
which we term as \textit{direct} ($\mathcal{V}_4$) and \textit{indirect} ($\mathcal{V}^{a-d}_3$).

The direct collision between the two main magnons is caused by the quartic vertex $\mathcal{V}_4$ while the cubic vertices $\mathcal{V}^{a-d}_3$ represent the
indirect magnon$-$magnon interactions. Note that in the direct ladder interaction events the two main magnons created in the RIXS process are stable while virtual decays and recombination are allowed in the indirect collision process.
Using Feynman rules in momentum space then yields the following equations for the two$-$particle propagator and the vertex function
\begin{align}\label{Eq:BSeq1}
\Pi_{\kk\kk'}(\qq,\omega)=&2i\int\frac{\mathrm{d}\omega'}{2\pi}\mathrm{G}_{\kk+\qq}(\omega+\omega')
\mathrm{G}_{-\kk}(-\omega')\Gamma_{\kk\kk'}(\omega,\omega'),\\ \label{Eq:BSeq2}
\Gamma_{\kk\kk'}(\omega,\omega')=&\delta_{\kk\kk'}+\sum_{\kk_1}2i\int\frac{\mathrm{d}\omega_1}{2\pi}
\mathrm{G}_{\kk_1+\qq}(\omega+\omega_1)\mathrm{G}_{-\kk_1}(-\omega_1)\nonumber\\
&\times\mathcal{V}^{\mathrm{IR}}_{\kk\kk_1}(\omega^\prime,\omega_1)\Gamma_{\kk_1\kk'}(\omega,\omega_1),
\end{align}
with the basic one-magnon propagator up to $1/S$ order defined as
\begin{eqnarray}
 \mathrm{G}^{-1}(\kk,\omega)&=&\omega-\omega_\kk+i0^+.
\end{eqnarray}
The factor of $2$ in Eq.(\ref{Eq:BSeq1}) and Eq.(\ref{Eq:BSeq2}) stem from
the two sets of contributions differing by the interchange of dummy momenta $\kk(\kk_{1})+\qq$ and $-\kk(\kk_{1})$ according to the Wick's theorem.
The lowest order two-particle irreducible interaction vertex $\mathcal{V}_{\mathrm{IR}}$, shown in Fig.~\ref{fig:bethe-salpeter}(c), reads as
\begin{eqnarray}
\mathcal{V}_{\mathrm{IR}}
=\mathcal{V}_4+\mathcal{V}_3^{(a)}+\mathcal{V}_3^{(b)}+\mathcal{V}_3^{(c)}+\mathcal{V}_3^{(d)},
\end{eqnarray}
where the frequency-independent four-point vertex $\mathcal{V}_4$ coming from the quartic Hamiltonian has the form
\begin{eqnarray}
\mathcal{V}_4=V_c(\kk_1+\qq,-\kk_1;\kk+\qq,-\kk),
\end{eqnarray}
and the other four vertices $\mathcal{V}_3^{(a-d)}$ in the same $1/S$ order which are assembled from two three-point vertices and one frequency-dependent
propagator can be written as
\begin{eqnarray}
\mathcal{V}_3^{(a)}=&&\frac{1}{(2!)^2}[V_{a}(\kk_1+\qq,\kk-\kk_1;\kk+\qq)\mathrm{G}_0(\kk-\kk_1,\omega'-\omega_1)\non\\
&&\times V^\ast_{a}(-\kk,\kk-\kk_1;-\kk_1)],\\
\mathcal{V}_3^{(b)}=&&\frac{1}{(2!)^2}[V^\ast_{a}(\kk+\qq,\kk_1-\kk;\kk_1+\qq)\mathrm{G}_0(\kk_1-\kk,\omega_1-\omega')\non\\
&&\times V_{a}(-\kk_1,\kk_1-\kk;-\kk)],\\
\mathcal{V}_3^{(c)}=&&\frac{1}{(2!)^2}[V_{a}(\kk_1+\qq,-\kk_1;\qq)\mathrm{G}_0(\qq,\omega)\non\\
&&\times V^\ast_{a}(\kk+\qq,-\kk;\qq)],\\
\mathcal{V}_3^{(d)}=&&\frac{1}{(3!)^2}[V_{b}(\kk_1+\qq,-\kk_1,\qq)\mathrm{G}_0(-\qq,-\omega)\non\\
&&\times V^\ast_{b}(\kk+\qq,-\kk,\qq)],
\end{eqnarray}
where we have retained only the bare propagator $\mathrm{G}_0$ for each intermediate line in $\mathcal{V}_3^{(a-d)}$ in the spirit of $1/S$ expansion. We further assume that two on-shell magnons are created and annihilated in the repeated ladder scattering process with $\omega'\approx-\varepsilon^{(0)}_\kk$ and $\omega_1\approx-\varepsilon^{(0)}_{\kk_1}$~\cite{PhysRevB.77.174412,PhysRevB.87.174423}.
This approximation is best for sharp spectral peaks of the two main magnons in the scattering process where all the lowest order irreducible vertices are not explicitly frequency dependent.
Based on the above simplifications, we now derive the final solution of the interacting RIXS intensity from the ladder approximation BS equation.

An approach to solving the coupled BS equations is to decompose the irreducible vertices into lattice
harmonics as demonstrated for the case of collinear antiferromagnet~\cite{PhysRevB.75.214414,PhysRevB.89.165103}.
An inspection of the interaction vertices for the TLAF reveals that $\mathcal{V}_3^{(a,b)}$ can not be
separated into finite sum of products of the triangular-lattice harmonics,
thus Eq.(\ref{Eq:corrfun}) can not be algebraically solved in terms of a finite number of scattering channels.
However, a numerical solution can be performed on finite lattices by summing over $N$ points of $\kk$ in the 1$^{st}$ BZ,
leading to a $N\times N$ system for the linear solver. We adopt this numerical approach to compute the interacting intensity plots.

We begin with substituting (\ref{Eq:BSeq1}) and (\ref{Eq:BSeq2}) into (\ref{Eq:corrfun}),
\begin{eqnarray}
\chi_2=\sum_{\kk\kk'}M_{\kk}M_{\kk'}
\Big[\delta_{\kk\kk^\prime}\Pi_{\kk}
+\Pi_{\kk}\sum_{\kk_1}V_{\kk\kk_1}\Pi_{\kk_1\kk'}\Big],
\end{eqnarray}
where $\Pi_{\kk}=2[\omega-\varepsilon_{\kk+\qq}-\varepsilon_\kk+i0^+]^{-1}$ represents the  renormalizated two-magnon propagator
in the absence of vertex correction.
The BZ on finite lattices can be divided into $\sqrt{N}\times \sqrt{N}$ meshes with the replacement of the continuous momenta $(\kk,\kk^\prime,\kk_1)$
to discrete variables $(\mathrm{m,n,l})$. The elements for the bimagnon susceptibility matrix  are given by
\begin{eqnarray}
\mathrm{\hat{\chi}}_{\mathrm{mn}}=M_\mathrm{m}M_\mathrm{n}\Big[\delta_\mathrm{mn}\Pi_\mathrm{m}
+\Pi_\mathrm{m}\sum_{\mathrm{l}}V_\mathrm{ml}\Pi_\mathrm{ln}\Big].
\end{eqnarray}
We then obtain the eigenvalue equation for these discrete momenta
\begin{eqnarray}\label{Eq:eigeneq}
\mathcal{A}_\mathrm{mn}&=&\mathcal{G}_\mathrm{mn}+\sum_{\mathrm{l}}\Gamma_\mathrm{ml}\mathcal{A}_\mathrm{ln},
\end{eqnarray}
where the new functions are defined as
\begin{eqnarray}
\mathcal{A}_\mathrm{mn}=\hat{\Pi}_\mathrm{mn}M_\mathrm{n}, ~\mathcal{G}_\mathrm{mn}=\delta_\mathrm{mn}\Pi_\mathrm{m}M_\mathrm{n},
~\Gamma_\mathrm{ml}=\Pi_\mathrm{m}V_\mathrm{ml}.
\end{eqnarray}
A direct solution to (\ref{Eq:eigeneq}) gives the final form of the $\mathrm{\hat{\chi}}$ matrix as
\begin{eqnarray}\label{Eq:fincorr2}
 \mathrm{\hat{\chi}}=\mathcal{\hat{D}}\textbf{[}\mathbf{\hat{1}}-\hat{\Gamma}\textbf{]}^{-1}\mathcal{\hat{G}},
\end{eqnarray}
where all the matrices in Eq. (\ref{Eq:fincorr2}) have $N\times N$ dimensions with the matrix elements explicitly defined as
\begin{eqnarray}
\mathbf{\hat{1}}_\mathrm{mn}&=&\delta_\mathrm{mn},\mathcal{\hat{D}}_\mathrm{mn}=\delta_\mathrm{mn}M_\mathrm{m},
\\\hat{\Gamma}_\mathrm{mn}&=&\Pi_\mathrm{m}V_\mathrm{mn},
~\mathcal{\hat{G}}_\mathrm{mn}=\delta_\mathrm{mn}\Pi_\mathrm{m}M_\mathrm{n}.
\end{eqnarray}
The interacting pure bimagnon RIXS susceptibility can then be computed as
\begin{eqnarray}\label{Eq:fincorr}
\mathrm{\chi}_{2}(\qq,\omega)=\sum_{\mathrm{m,n}}\mathrm{\hat{\chi}}_{\mathrm{mn}}.
\end{eqnarray}

\begin{figure*}
\centering\includegraphics[scale=0.5]{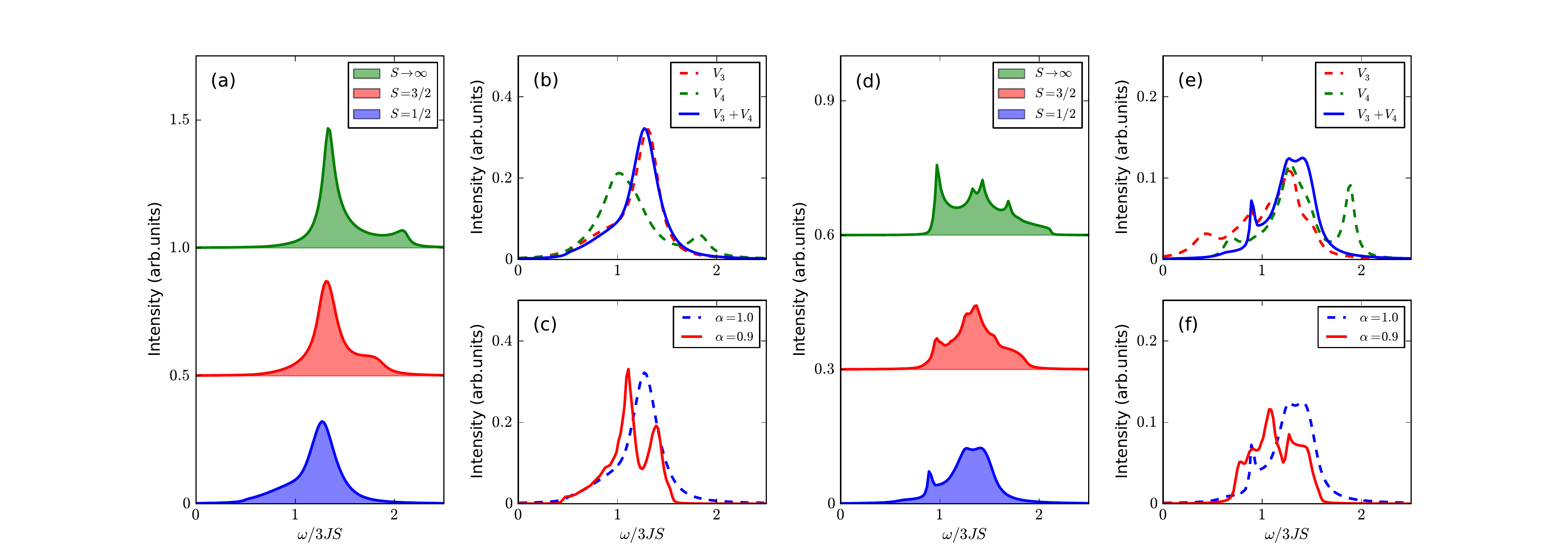}
\caption{(Color online)
Interacting bimagnon RIXS intensity for transformed momenta $\qq=M$ (a-c) and $\qq=Y$ (d-f).
In (a) and (d) the evolution of the interacting bimagnon intensity profile with increasing spin value $S$ for the isotropic model is shown.
In (a) and (d): $S \rightarrow \infty$ (top frame), $S=3/2$ (middle frame), and $S=1/2$ (bottom frame). In (b) and (e) comparison of
the contribution from the direct ($\mathcal{V}_4$), indirect ($\mathcal{V}_3$) and
full ($\mathcal{V}_3+\mathcal{V}_4$) vertices correction to the RIXS spectrum for $S=1/2$ and $\alpha=1$ are displayed.
In (c) and (f) the effects of easy-plane anisotropy on the splitting feature of the interacting bimagnon spectrum are shown.
}
\label{fig:rixsgm}
\end{figure*}

In Fig.~\ref{fig:rixsgm} we plot the results for interacting bimagnon RIXS intensity. We choose two special BZ momenta values, $M$ and $Y$, to illustrate our findings. In Fig.~\ref{fig:rixsgm}(a) and Fig.~\ref{fig:rixsgm}(d) we show the progression of the indirect RIXS spectra shape as the spin $S$ value is changed from the classical case $S \rightarrow \infty$ (top), to $S = 3/2$ (middle), to the maximal quantum case of $S = 1/2$ (bottom). While the classical RIXS spectra from both momenta contain peaks due to the presence of Van Hove singularities, introduction of quantum fluctuations cause some of these spurious peaks to disappear. But observe that in the S = $1/2$ case the spectra shape is strikingly different. In the absence of anisotropy at the $\qq=M$ (roton transfer momentum), we observe a single peak at an energy of $4JS$. However, at the $\qq=Y$ point there is a multipeak structure, see Fig.~\ref{fig:rixsgm}(d). Now comparing with the magnon decay intensity map, Fig.~\ref{fig:sw}(b), it is evident that there is a direct correlation between the stability of the spin wave modes and the appearance of a single or multipeak structure. The above mentioned comparison is not restricted to these two choosen points. Comparision of the RIXS spectra generated from other special high symmetry momentum transfer also have the same features, see Fig.~\ref{fig:fullspectra}. Based on these observations we propose that RIXS can be used as a probe to detect the presence of the roton mode in a TLAF.  Furthermore, to provide a comprehensive picture of the effects of geometrical frustration and anisotropy we introduce a small anisotropy $\alpha=0.9$ in the system. From Fig.~\ref{fig:rixsgm}(c) and Fig.~\ref{fig:rixsgm}(f) it is clear that inclusion of anisotropy causes further peak splitting. Thus a proper explanation of the RIXS spectra features in a TLAF involves analyzing both the effects of magnon damping and anisotropy.

The $\mathcal{V}_4$ and $\mathcal{V}_3$ vertices play an important role in the generation of the RIXS
bimagnon spectra. Especially at the roton point it is worth noting that including only the direct collision vertex does not renormalize the single$-$peak structure in the extreme quantum condition with $S=1/2$, Fig.~\ref{fig:rixsgm}(b). The major contribution to the interacting RIXS spectra originates from the indirect vertices arising from the three-magnon interaction terms. This indicates that renormalization of the spectra is due to the \emph{indirect} $\mathcal{V}_3$ vertices which involve virtual decay and recombination of the two main magnons in the scattering process. This is different from the $\qq=Y$ point where both the $\mathcal{V}_4$ and $\mathcal{V}_3$ vertices contribute, as seen in Fig.~\ref{fig:rixsgm}(e).

\begin{figure}
\includegraphics[scale=0.3]{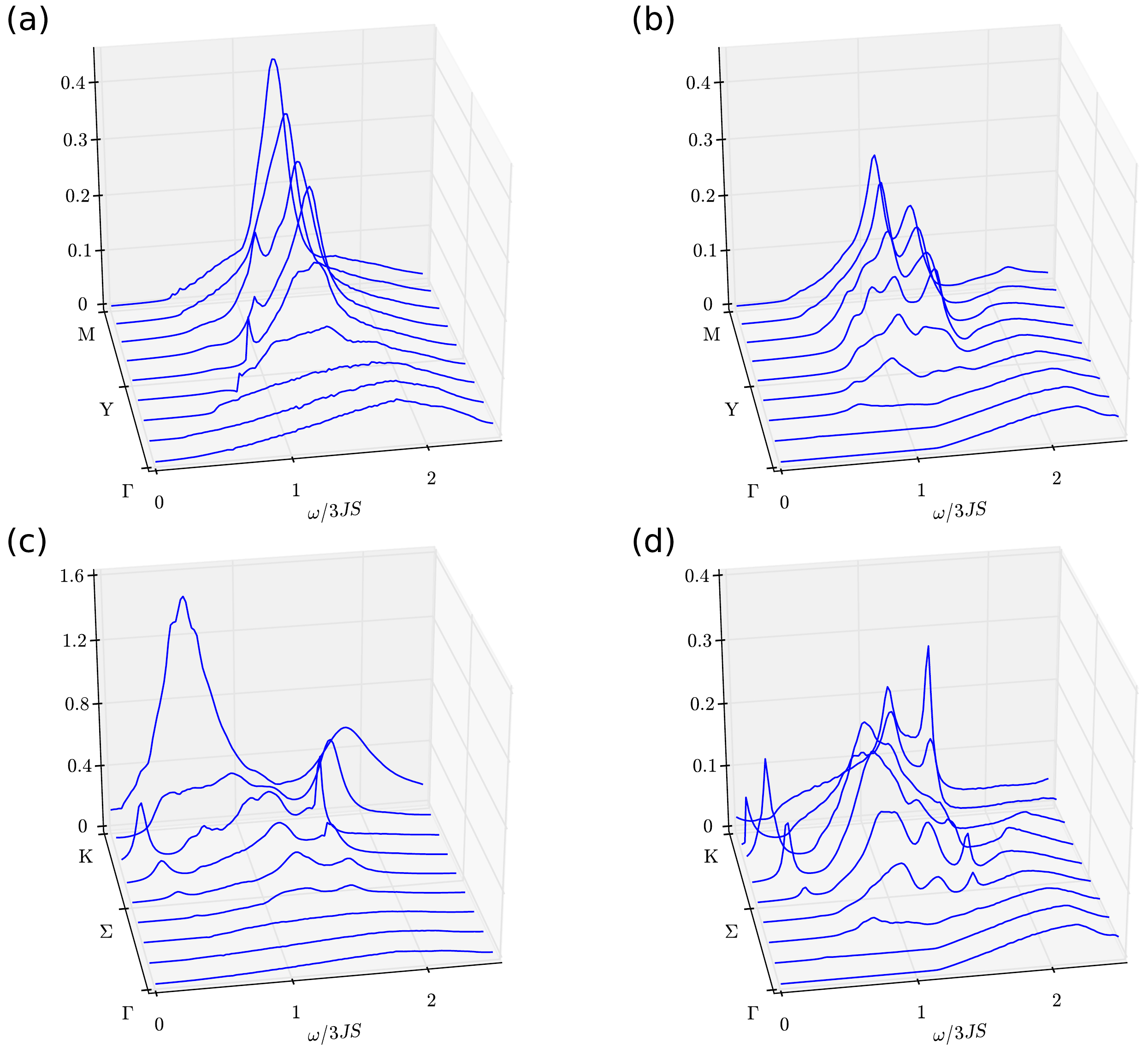}
\caption{(Color online)
Full indirect RIXS spectra $I_2+I_3$ of the $S=1/2$ $XXZ$ triangular$-$lattice antiferromagnet with $\alpha=1$; (a) and (c) and $\alpha=0.9$; (b) and (d),  along the high symmetry path $\Gamma\rightarrow M$ and  $\Gamma\rightarrow K$ in the BZ.}
\label{fig:fullspectra}
\end{figure}

Before we end this section it is important to point out an important difference between a Raman scattering calculation
and RIXS. In the case of RIXS, the contributions of diagrams $\mathcal{V}_3^{(c)}$ and $\mathcal{V}_3^{(d)}$ vanish identically when the transferred momenta belongs to the $\Gamma - M$ path or related symmetrical lines in the BZ in accordance with the magnetic Raman scattering study for which $\qq=0$~\cite{PhysRevB.77.174412}.
To demonstrate this fact we consider the contributions of the two$-$particle irreducible vertices $\mathcal{V}_3^{(c)}$ and $\mathcal{V}_3^{(d)}$ which are already
in separated forms as functions of $\kk$ and $\kk^\prime$. The corresponding reducible vertex function Eq.~(\ref{Eq:BSeq2})
with respect to these diagrams can be directly obtained as
\begin{align}
\Gamma_{\kk\kk'}=&\delta_{\kk\kk'}+V^\ast_{a(b)}(\kk)f(\qq,\omega),
\end{align}
where $f(\qq,\omega)$ is a function of $\omega$ and $\qq$ only. Barring the noninteracting contributions the vertex correction to the RIXS susceptibility is given by
\begin{align}
\chi_{V}(\qq,\omega)&=f(\qq,\omega)\sum_{\kk^\prime}M_{\kk^\prime}\sum_{\kk}M_\kk\Pi_{\kk}V^\ast_{a(b)}(\kk)\non\\
&=\mathrm{const}\times\sum_{\kk}M_\kk\Pi_{\kk}V^\ast_{a(b)}(\kk).
\end{align}
In the above both $M_\kk$ and $\Pi_{\kk}$ are even functions of $\kk$, while the function $V^\ast_{a(b)}(\kk)=V^\ast_{a(b)}(\kk+\qq,-\kk,\qq)$ are odd functions with respect
to $\kk$ when momentum $\qq$ is along the $\Gamma M$ line (e.g. $q_x=0$).
Thus, by virtue of the $C_3$ symmetry of hexagonal lattices we can conclude that the total contributions of diagrams $\mathcal{V}_3^{(c)}$ and $\mathcal{V}_3^{(d)}$ vanish identically
when transformed momenta $\qq$ are located in the lines from the center of the BZ to the middle of the BZ boundary.
This implies that the source processes are prohibited in the repeated ladder scattering events when transferred momenta are along these symmetrical paths.

\subsection{Total RIXS intensity}\label{Sec:Total}
\begin{figure}
\includegraphics[scale=0.3]{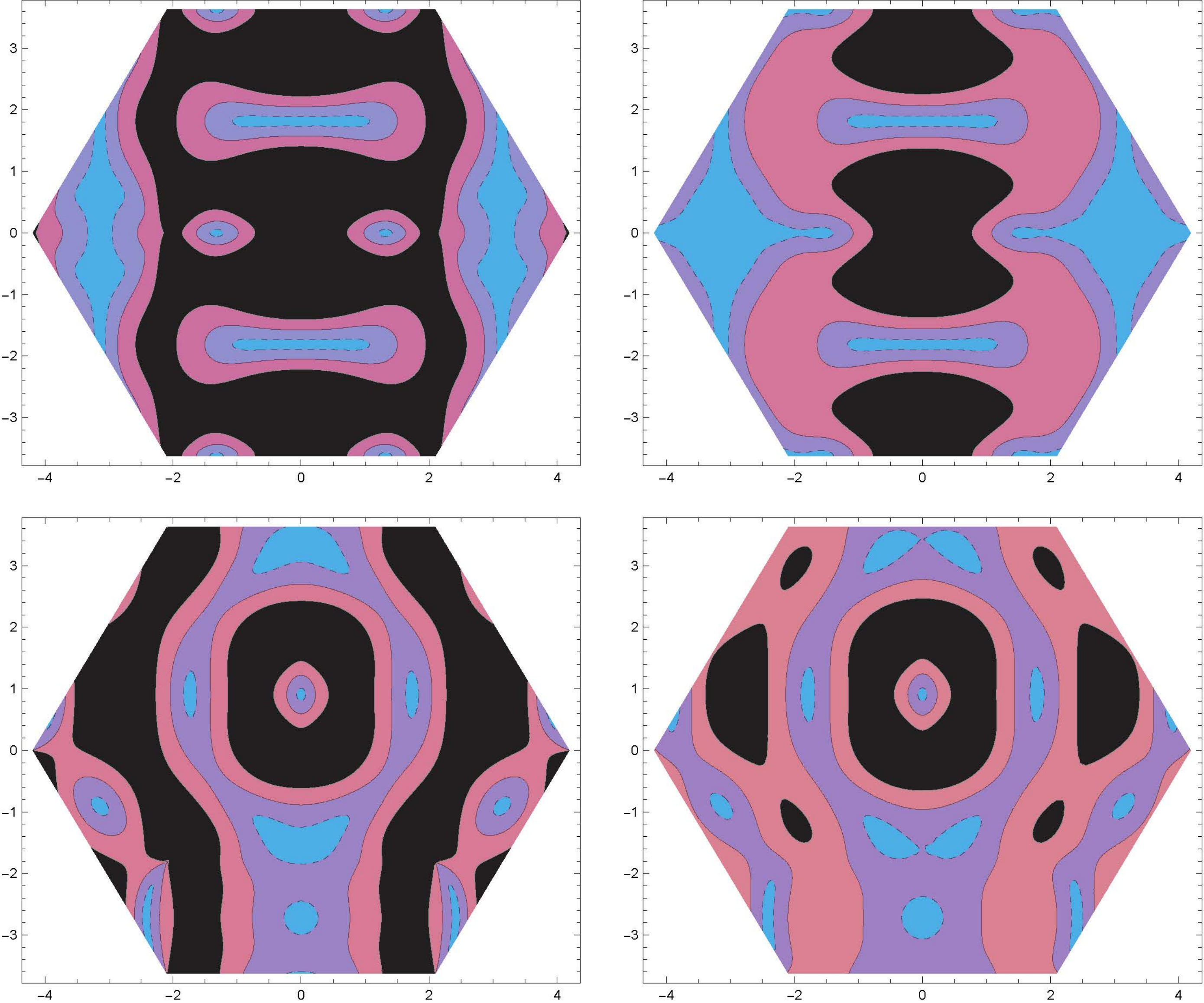}
\caption{(Color online)
Contour plot of the bimagnon velocity in the first BZ of TLAF for transferred momenta $\qq=M$ (first row) and $\qq=Y$ (second row) with $\alpha=1$ and $\alpha=0.9$. First column is the bimagnon velocity in the absence of anisotropy. Second column is for the anisotropic ($\alpha=0.9$) case. Lowest velocity contours indicated by blue (grey) bands. Relative higher velocities indicated by black regions.
}
\label{fig:velgraph}
\end{figure}
\begin{figure*}
\includegraphics[scale=0.5]{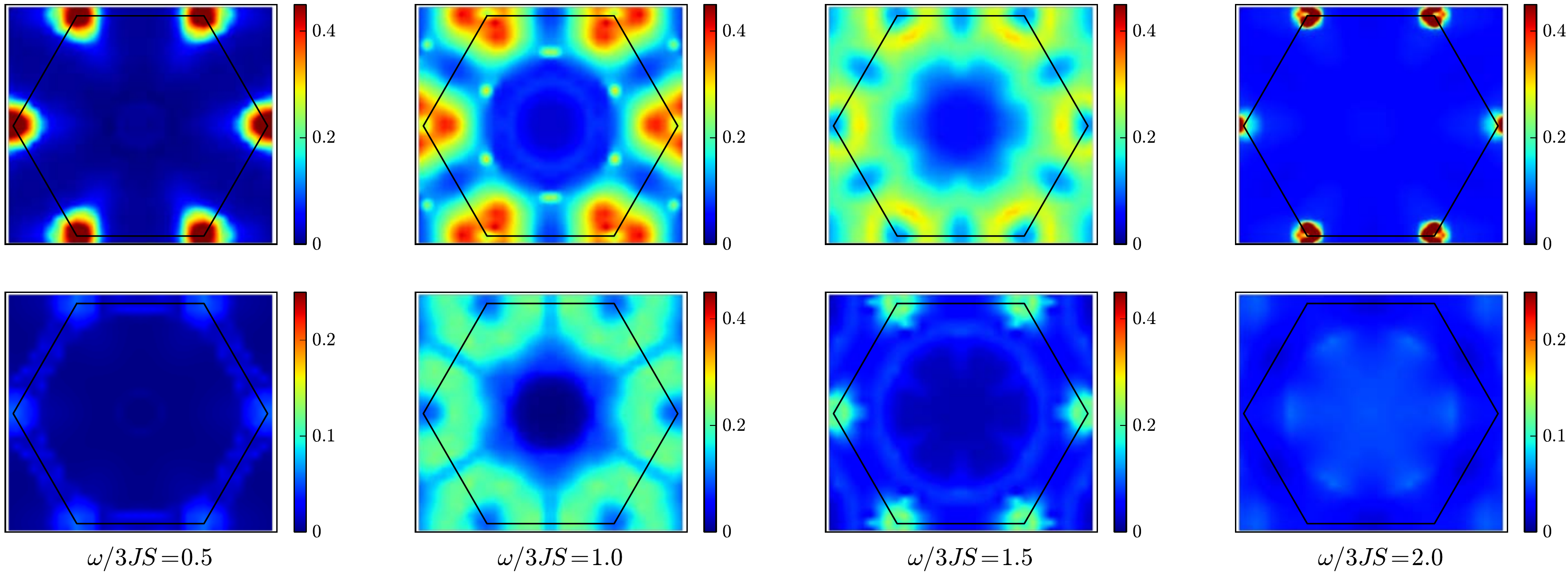}
\caption{(Color online)
Intensity plots of the constant-$\omega$ scans of the full RIXS intensity $I(\qq,\omega)$ scaled by $(3JS)^2$ for the $S=1/2$ $XXZ$ triangular lattice antiferromagnet
in the $\qq$ plane with $\alpha=1$ (upper) and $\alpha=0.9$ (lower) at four representative energies. The prominent peaks on the corners are strongly reduced in the presence
of anisotropy in qualitative agreement with the noninteracting total spectral weight.
}
\label{fig:omegascan}
\end{figure*}
Using Eq.~(\ref{fullrixs}) we compute the full indirect RIXS spectra. In Fig.~\ref{fig:fullspectra} we display
the RIXS line plots along the $\Gamma\rightarrow M$ path and along the $\Gamma\rightarrow K$ path, respectively. The features
observed are reminiscent of those discussed for the noninteracting trimagnon spectra and the full interacting bimagnon spectra. As noted earlier, we find that at the $\Gamma$ point the spectra originates purely from the trimagnon contribution, irrespective of the presence or absence of anisotropy. However, inclusion of anisotropy causes a downshift of the bimagnon contribution and an upward shift of the trimagnon spectra. Anisotropy gives rise to further splitting in the bimagnon case, however, the trimagnon spectra is not affected. The occurence of peak splitting observed in the RIXS spectrum can be predicted by observing the bimagnon velocity plot. In a previous publication on the square lattice Heisenberg magnet~\cite{PhysRevB.89.165103} we had highlighted the connection between bimagnon velocity and the appearance of multipeak structure in the RIXS spectra. Interestingly enough, even within the context of a TLAF this relationship persists. To demonstrate this correlation, in Fig.~\ref{fig:velgraph}, we show the bimagnon velocity intensity plot in both the presence and absence of anisotropy for the $\qq=M$ and the $\qq=Y$ point. The black regions represent the highest moving bimagnon velocities which clearly disappear with the inclusion of anistropy. As more puddles of slow moving bimagnon velocity appears, so does the appearance of a multipeak structure as shown in Fig.~\ref{fig:fullspectra}. At the $\qq=M$ point the single roton peak melts away with increasing anistropy which comes along with low bimagnon velocity. A similar effect is observed at the $\qq=Y$ point, where with increasing anisotropy there are greater pockets of slow moving bimagnon. Hence, with anisotropy the peak splits further at $Y$ point. Finally, in Fig.~\ref{fig:omegascan} we present the expected constant$-\omega$ scans of the total interacting RIXS intensity for four selected energies from low energy to a high energy. One of the advantages of these constant$-$energy scans, which is reminiscent of the INS experiments, is that prominent peak structures are easy to distinguish~\cite{PhysRevB.88.094407}.
For the isotropic model the interacting RIXS intensity is strongly peaked on the corners of the hexagonal BZ at low and high energies, while these peaks disperse along the edges of the BZ at
intermediate energies. However the presence of $XXZ$ anisotropic strongly reduces these prominent features,
which is in qualitative agreement with the noninteracting total spectral weight shown in Fig.~\ref{fig:rixsbare}.

\section{Conclusion}\label{Sec:Conclu}
At present, there exists no theoretical guidance for experimentalists on how to analyze and
interpret the RIXS spectra of an ordered phase in a geometrically frustrated quantum magnet.
Although a proposal for detecting spin$-$chirality terms in triangular lattice Mott insulators via RIXS
has been put forward~\cite{winghoPhysRevB.84.125102}, there has been no analysis on the effect of geometrical frustration and anisotropy on the indirect RIXS spectra. In this paper, using a $1/S$ expansion spin$-$wave theory involving Bethe$-$Salpeter corrections we investigate the key signatures of noncollinear ground state ordering in the indirect RIXS spectrum of a TLAF. We conclude that in the absence of anisotropy the root cause of the multipeak structure is magnon decay. This mechanism is different from that of a square lattice where strong frustrating further neighbor interactions and
anisotropy are required to cause peak splitting (instability). In the introduction we had put forward a couple of questions$-$ (a) How does the presence of an intrinsic damping affect the indirect K$-$edge RIXS spectra? and (b) What role does the interplay between geometrical frustration and spin anisotropy have on the RIXS spectra? Based on our calculations, we conclude that magnon damping does affect the spectra, causing the RIXS peak to be either stable (no splitting) or unstable (splitting leading to multipeak) in the absence or presence of damping, respectively. Geometrical frustration introduces noncollinear ordering which introduces magnon damping. The stability or instability of the ensuing magnon mode then dictates the appearance of a single or multipeak structure. By comparing the K$-$edge RIXS intensity of the square lattice case, to that of the TLAF, we find that  the RIXS intensity does not vanish at the $\Gamma$ point and at the antiferromagnetic wavevector. At the $\Gamma$ point, the bimagnon intensity is zero and the single peak spectra results purely from the trimagnon contribution, approximately at energy scale of $6JS$ corresponding to the three magnon energy. This provides experimentalists with a means to detect purely trimagnon RIXS spectra at the K$-$edge. Our proposed scheme of detecting trimagnon excitations is different from that put forward in the paper by Ament and Brink~\cite{arXiv:1002.3773}, since we are not considering the L$-$edge. The single roton peak occurs at an energy of $4JS$ and can be used as an experimental signature to detect roton modes in a TLAF. In conclusion, our theoretical investigation demonstrates that RIXS has the potential to probe and provide a comprehensive characterization of the microscopic properties of bimagnon and trimagnon excitations in the TLAF across the entire BZ, which is beyond the capabilities of traditional low$-$energy optical techniques~\cite{RevModPhys.79.175,JPCM.19.145243,PhysRevB.77.174412,PhysRevB.87.174423}.

\begin{acknowledgments}
T.D. acknowledges funding support from Cottrell Research Corporation grant and Georgia Regents University Small Grants program.
C.L., Z.H., and D.X.Y. acknowledge support from National Basic Research Program of China (2012CB821400), NSFC-11074310, NSFC-11275279, RFDPHE-20110171110026, NCET-11-0547,
and Fundamental Research Funds for the Central Universities of China.
\end{acknowledgments}

\appendix

\section{Derivation of interacting spin-wave theory}\label{App:SWT}
We utilize the Holstein-Primakoff transformation to bosonize the
local rotating Hamiltonian (\ref{localham})
\begin{eqnarray}
 S_i^z=S-a_i^\dag a_i,\ S_i^-=a^\dag\sqrt{2S-a_i^\dag a_i},\ S_i^+=(S_i^-)^\dag,
\end{eqnarray}
with subsequent expansion of square root to first order in $a_i^\dag a_i/2S$.
This is followed by a Fourier transformation. The Fourier transformed Hamiltonian takes the form
\begin{eqnarray}
   \mathcal{H}=H_0+H_2+H_3+H_4+O(S^{-1}).
\end{eqnarray}
The first term corresponds to the classical energy and the quadratic Hamiltonian reads
\begin{eqnarray}
  &&H_2=\sum_\kk \Big[A_\kk a_\kk^\dag a_\kk+\frac{1}{2}B_\kk(a_\kk^\dag a_{-\kk}^\dag+a_{-\kk}a_\kk)\Big],\non\\
  &&A_\kk=3JS[1+(\alpha-\frac{1}{2})\gamma_\kk],\ B_\kk=-3JS(\alpha+\frac{1}{2})\gamma_\kk,
\end{eqnarray}
with the structure factor $\gamma_\kk$ defined as
\begin{eqnarray}
  \gamma_\kk=\frac{1}{3}\Big(\cos k_x+2\cos\frac{k_x}{2}\cos\frac{\sqrt{3}}{2}k_y\Big).
\end{eqnarray}
We then diagonalize the harmonic part $H_2$ by the Bogoliubov transformation
\begin{eqnarray}
 a_\kk=u_\kk b_\kk+v_\kk b_{-\kk}^\dag,
\end{eqnarray}
with the parameters $u_{\kk}$ and $v_{\kk}$ defined as
\begin{eqnarray}\label{Eq.bgpara}
u_{\kk}^2,v_{\kk}^2=\frac{A_\kk\pm\varepsilon_\kk}{2\varepsilon_\kk},
u_{\kk}v_{\kk}=-\frac{B_\kk}{2\varepsilon_\kk},
\end{eqnarray}
and the linear spin-wave theory dispersion given by
\begin{eqnarray}
\varepsilon_\kk=\sqrt{A_\kk^2-B_\kk^2}=3JS\epsilon_\kk,
\end{eqnarray}
where we have defined the dimensionless energy
\begin{eqnarray}
  \epsilon_\kk=\sqrt{(1-\gamma_{\kk})(1+2\gamma_\kk)}.
\end{eqnarray}

Performing the Bogoliubov transformations in the cubic interaction term $H_3$ we obtain
\begin{eqnarray}
   H_3=&&\frac{1}{2!}\sum_{\kk_1+\kk_2=\kk_3}V_{a}(\kk_1,\kk_2;\kk_3)(b_{\kk_1}^\dag b_{\kk_2}^\dag b_{\kk_3}+\mathrm{H.c.})\non\\
 &&+\frac{1}{3!}\sum_{\kk_1+\kk_2+\kk_3=0}V_{b}(\kk_1,\kk_2,\kk_3)(b_{\kk_1}^\dag b_{\kk_2}^\dag b_{\kk_3}^\dag+\mathrm{H.c.})
\end{eqnarray}
The explicit forms for the three-boson interaction vertices are
\begin{eqnarray}
  V_{a}(1,2;3)=&&3Ji\sqrt{\frac{3S}{2}}\big[\bar{\gamma}_1(u_1+v_1)(u_2u_3+v_2v_3)+\bar{\gamma}_2(u_2+v_2)\non\\
  &&\times(u_1u_3+v_1v_3)-\bar{\gamma}_3(u_3+v_3)(u_1v_2+v_1u_2)\big],\\
  V_{b}(1,2,3)=&&3Ji\sqrt{\frac{3S}{2}}\big[\bar{\gamma}_1(u_1+v_1)(u_2v_3+v_2u_3)+\bar{\gamma}_2(u_2+v_2)\non\\
  &&\times(u_1v_3+v_1u_3)+\bar{\gamma}_3(u_3+v_3)(u_1v_2+v_1u_2)\big],
\end{eqnarray}
where $u_i$, $v_i$ are Bogoliubov parameters and the function $\bar{\gamma}_\kk$ is defined as
\begin{eqnarray}
  \bar{\gamma}_\kk=&&\frac{1}{3}\Big(\sin k_x-2\sin\frac{k_x}{2}\cos\frac{\sqrt{3}}{2}k_y\Big).
\end{eqnarray}
The three-boson vertex $V_{a}$ and $V_{b}$ in $H_3$ describes interaction between one- and two-magnon states and are called the decay and
the source vertex, respectively.

To derive the explicit forms of the quartic interaction term $H_4$, it is convenient to introduce the following Hartree-Fock averages
\begin{eqnarray}
  n&=&\langle a_i^\dag a_i\rangle=\frac{1}{2}c_0+\frac{2\alpha-1}{4}c_1-\frac{1}{2},\\
  m&=&\langle a_i^\dag a_j\rangle=\frac{1}{2}c_1+\frac{2\alpha-1}{4}c_2,\\
  \Delta&=&\langle a_i a_j\rangle=\frac{2\alpha+1}{4}c_2,\\
  \delta&=&\langle a_i^2\rangle=\frac{2\alpha+1}{4}c_1,
\end{eqnarray}
with the two-dimensional integrals
\begin{eqnarray}
  c_l=\sum_\kk\frac{(\gamma_\kk)^l}{\epsilon_\kk},\ (l=0,1,2).
\end{eqnarray}
After the mean-field decoupling, the quartic part is decomposed as
\begin{eqnarray}
  H_4=\delta H_0+\delta H_2+\tilde{H}_4.
\end{eqnarray}
The first term $\delta H_0$ is the correction to the ground-state energy and the quadratic parts reads
\begin{eqnarray}
  \delta H_2=\sum_\kk \Big[\delta A_\kk a_\kk^\dag a_\kk+\frac{1}{2}\delta B_\kk(a_\kk^\dag a_{-\kk}^\dag+a_{-\kk}a_\kk)\Big],
\end{eqnarray}
with
\begin{eqnarray}
  \delta A_\kk=&&\frac{3}{2}[(1+2\alpha)\Delta+(1-2\alpha)m-2n]\non\\
  &&+\frac{3}{4}[(1+2\alpha)\delta+2(1-2\alpha)n-4m]\gamma_\kk,\non\\
  \delta B_\kk=&&\frac{3}{4}[(1+2\alpha)m+(1-2\alpha)\Delta]\non\\
  &&+\frac{3}{4}[(1-2\alpha)\delta+2(+2\alpha)n-4\Delta]\gamma_\kk.
\end{eqnarray}
We then obtain the Hartree-Fock correction to the harmonic spin-wave spectrum
\begin{eqnarray}
 \delta\varepsilon_\kk=(u_\kk^2+v_\kk^2)\delta A_\kk+2u_\kk v_\kk\delta B_\kk.
\end{eqnarray}
The normal-ordered term $\tilde{H}_4$ describes the multi-particle interactions. Here we only display the explicit expression for the lowest order irreducible
two-particle scattering amplitude which is relevant for the our calculations as
\begin{eqnarray}
     \tilde{H}_4^{2-p}=\sum_{\kk_1+\kk_2=\kk_3+\kk_4}V_{c}(\kk_1,\kk_2;\kk_3,\kk_4)b_{\kk_1}^\dag b_{\kk_2}^\dag b_{\kk_3}b_{\kk_4},
\end{eqnarray}
with the vertex function
\begin{widetext}
\begin{eqnarray}
  V_{c}(1,2;3,4)=&&\frac{3JS}{16S}\Big\{(2\alpha+1)(\gamma_1+\gamma_2+\gamma_4)(u_1u_2u_3v_4+v_1v_2v_3u_4)
                 +(2\alpha+1)(\gamma_1+\gamma_2+\gamma_3)(u_1u_2v_3u_4+v_1v_2u_3v_4)\non\\
                 &&+(2\alpha+1)(\gamma_2+\gamma_3+\gamma_4)(u_1v_2u_3u_4+v_1u_2v_3v_4)
                 +(2\alpha+1)(\gamma_1+\gamma_3+\gamma_4)(u_1v_2v_3v_4+v_1u_2u_3u_4)\non\\
                 &&-[2(\gamma_{1-3}+\gamma_{2-3}+\gamma_{1-4}+\gamma_{2-4})
                 +(2\alpha-1)(\gamma_1+\gamma_2+\gamma_3+\gamma_4)](u_1u_2u_3u_4+v_1v_2v_3v_4)\non\\
                 &&-[2(\gamma_{1+2}+\gamma_{3+4}+\gamma_{1-3}+\gamma_{2-4})
                 +(2\alpha-1)(\gamma_1+\gamma_2+\gamma_3+\gamma_4)](u_1v_2u_3v_4+v_1u_2v_3u_4)\non\\
                 &&-[2(\gamma_{1+2}+\gamma_{3+4}+\gamma_{1-4}+\gamma_{2-3})
                 +(2\alpha-1)(\gamma_1+\gamma_2+\gamma_3+\gamma_4)](u_1v_2v_3u_4+v_1u_2u_3v_4)\Big\}.
\end{eqnarray}
\end{widetext}
By collecting all these terms together, we finally obtain the effective interacting spin$-$wave Hamiltonian Eq.~(\ref{effectiveham}).

\section{Exact versus numerical solution to the BS equation}\label{Exact solution}
\begin{table}
\caption{\label{Table:Channel}
Definition of the channels $v_\mathrm{n}(\kk)$
}
\begin{ruledtabular}
\begin{tabular}{cc|cc}
$n$ &  $v_{n}(\kk)$                 & $n$ & $v_{n}(\kk)$\\ \hline
 1 & $u_{\kk+\qq} u_{\kk}$          & 15  & $u_{\kk+\qq} v_{\kk} \cos k_x\cos \frac{\sqrt{3}}{2}k_y$\\
 2 & $v_{\kk+\qq} v_{\kk}$          & 16  & $v_{\kk+\qq} u_{\kk} \cos k_x\cos \frac{\sqrt{3}}{2}k_y$\\
 3 & $u_{\kk+\qq} v_{\kk}$          & 17  & $u_{\kk+\qq} u_{\kk} \sin k_x\sin \frac{\sqrt{3}}{2}k_y$\\
 4 & $v_{\kk+\qq} u_{\kk}$          & 18  & $v_{\kk+\qq} v_{\kk} \sin k_x\sin \frac{\sqrt{3}}{2}k_y$\\
 5 & $u_{\kk+\qq} u_{\kk} \cos k_x$ & 19  & $u_{\kk+\qq} v_{\kk} \sin k_x\sin \frac{\sqrt{3}}{2}k_y$\\
 6 & $v_{\kk+\qq} v_{\kk} \cos k_x$ & 20  & $v_{\kk+\qq} u_{\kk} \sin k_x\sin \frac{\sqrt{3}}{2}k_y$\\
 7 & $u_{\kk+\qq} v_{\kk} \cos k_x$ & 21  & $u_{\kk+\qq} u_{\kk} \cos k_x\sin \frac{\sqrt{3}}{2}k_y$\\
 8 & $v_{\kk+\qq} u_{\kk} \cos k_x$ & 22  & $v_{\kk+\qq} v_{\kk} \cos k_x\sin \frac{\sqrt{3}}{2}k_y$\\
 9 & $u_{\kk+\qq} u_{\kk} \sin k_x$ & 24  & $u_{\kk+\qq} v_{\kk} \cos k_x\sin \frac{\sqrt{3}}{2}k_y$\\
10 & $v_{\kk+\qq} v_{\kk} \sin k_x$ & 24  & $v_{\kk+\qq} u_{\kk} \cos k_x\sin \frac{\sqrt{3}}{2}k_y$\\
11 & $u_{\kk+\qq} v_{\kk} \sin k_x$ & 25  & $u_{\kk+\qq} u_{\kk} \sin k_x\cos \frac{\sqrt{3}}{2}k_y$\\
12 & $v_{\kk+\qq} u_{\kk} \sin k_x$ & 26  & $v_{\kk+\qq} v_{\kk} \sin k_x\cos \frac{\sqrt{3}}{2}k_y$\\
13 & $u_{\kk+\qq} u_{\kk} \cos k_x\cos \frac{\sqrt{3}}{2}k_y$ & 27  & $u_{\kk+\qq} v_{\kk} \sin k_x\cos \frac{\sqrt{3}}{2}k_y$\\
14 & $v_{\kk+\qq} v_{\kk} \cos k_x\cos \frac{\sqrt{3}}{2}k_y$ & 28  & $v_{\kk+\qq} u_{\kk} \sin k_x\cos \frac{\sqrt{3}}{2}k_y$\\
\end{tabular}
\end{ruledtabular}
\end{table}
\begin{figure}
\centering\includegraphics[scale=0.5]{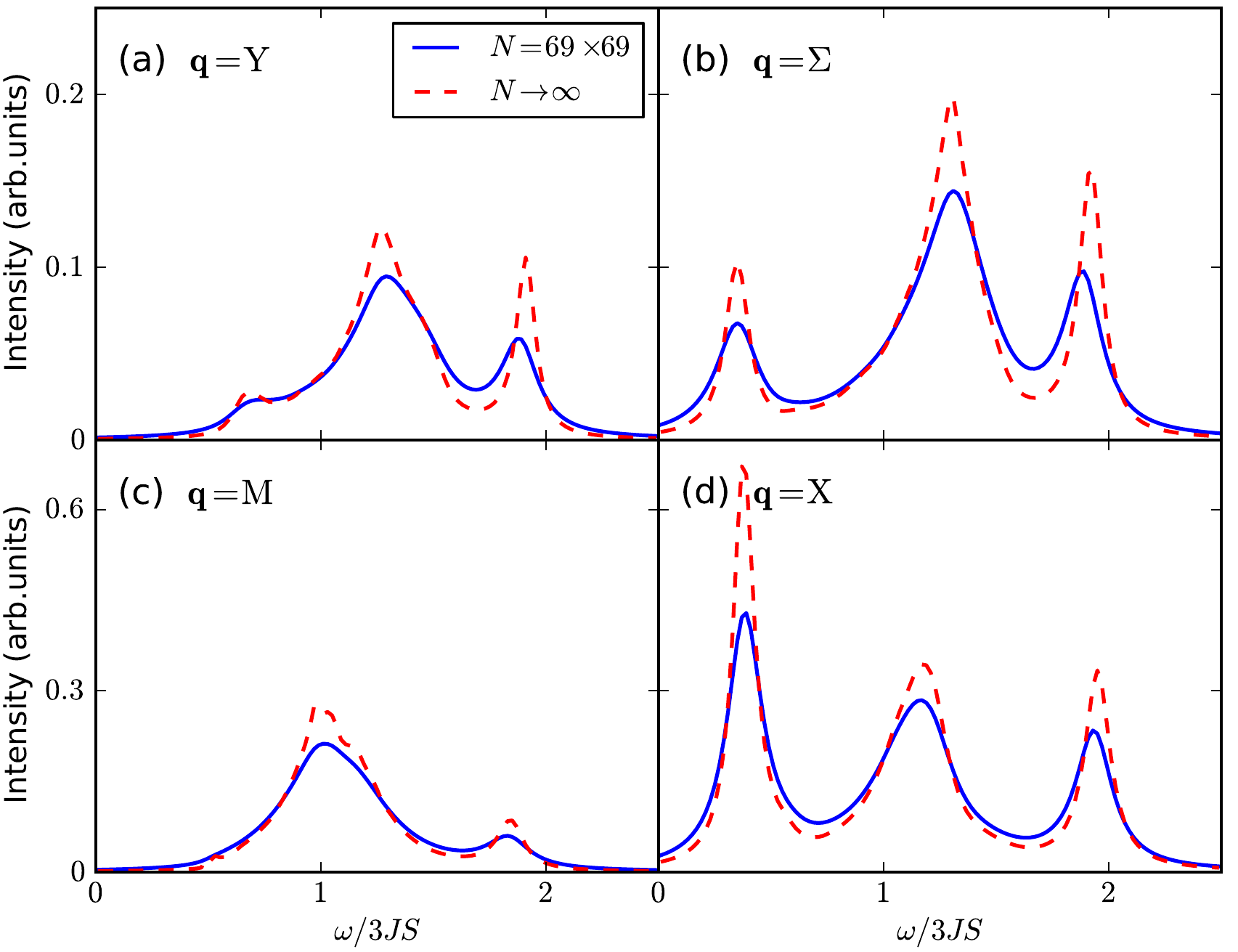}
\caption{(Color online)
Interacting bimagnon RIXS intensity renormalized only by the direct interaction vertex ($\mathcal{V}_4$) based on the
exact ($N\rightarrow\infty$) versus numerical ($N=69\times69$) solution of the Bethe-Salpeter equation
for a $S=1/2$ isotropic triangular lattice antiferromagnet at various momenta in the BZ.
}
\label{fig:rixsv4}
\end{figure}
To test the validity of our numerical method on finite lattices ($N=69\times69$), we adopt the exact solution approach to solving a BS equation
outlined in Appendix B of our publication Ref.~[\onlinecite{PhysRevB.89.165103}]. We obtain a separated form for the four-point vertex $\mathcal{V}_4$ for the Heisenberg model ($\alpha=1$) on triangular lattice which has the following expression
\begin{eqnarray}
\mathcal{V}_4(\kk_1+\qq,-\kk_1;\kk+\qq,-\kk)=\sum_{\mathrm{m,n}=1}^{\mathrm{28}} v_{\mathrm{m}}(\kk) \hat{\Gamma}_{\mathrm{m n}} v_{\mathrm{n}}(\kk_1).
\end{eqnarray}
The channels $v_\mathrm{n}(\kk)$ are defined in Table \ref{Table:Channel} with the matrix elements of $\hat{\Gamma}$ denoted by
\begin{equation}
\hat{\Gamma}=\frac{3JS}{16S}
\left(
\begin{array}{cccc}
\hat{S}_1 & \hat{T} \\
\hat{T} & \hat{S}_2 \\
\end{array}
\right),
\end{equation}
where the blocks are given by
\begin{widetext}
\begin{equation*}
\hat{T}=
\left(
\begin{array}{cccccccccccccc}
2 & 2\lambda  & -\frac{2}{3}\phi  & -\frac{2}{3}\phi  & 0 & 0 &  0  &  0  &  0  &  0  &  0  &  0  &  0  &  1    \\
2\lambda   & 2 & -\frac{2}{3}\phi  & -\frac{2}{3}\phi  &  0 & 0 &  0   & 0  &  0 &  0 &  0 & 0  &  0 & 2  \\
-\frac{2}{3}\chi   & 0 & 2\chi & 2\chi  & 0 &  0 &  0 & 0  & 0 & 0 & 0  &  0 & -\frac{8}{3}\chi  & 0 \\
0   & -\frac{2}{3}\chi & 2\chi & 2\chi  & 0  & 0   & 0 & 0 &0  & 0 & 0  & 0 &  0 & -\frac{8}{3}\chi    \\
0  & 2\chi & -\frac{2}{3}\chi  & -\frac{2}{3}\chi  & 0 & 0 & 0 & 0 & 0 & 0 & 0  & 0 & 0 & 0   \\
2\chi  & 0 & -\frac{2}{3}\chi  & -\frac{2}{3}\chi  & 0 & 0 & 0 & 0 & 0 & 0 & 0  & 0 & 0 & 0   \\
\frac{2}{3}\mu  & 0 & -2\mu  & -2\mu  & 0 & 0 & 0 & 0 & 0 & 0 & 0  & 0 & \frac{8}{3}\mu & 0   \\
0  & \frac{2}{3}\mu & -2\mu  & -2\mu  & 0 & 0 & 0 & 0 & 0 & 0 & 0  & 0 & 0 & \frac{8}{3}\mu   \\
0  & -2\mu & \frac{2}{3}\mu  & \frac{2}{3}\mu  & 0 & 0 & 0 & 0 & 0 & 0 & 0  & 0 & 0 & 0   \\
-2\mu  & 0 & \frac{2}{3}\mu  & \frac{2}{3}\mu  & 0 & 0 & 0 & 0 & 0 & 0 & 0  & 0 & 0 & 0   \\
\frac{2}{3}\nu  & 0 & -2\nu  & -2\nu  & 0 & 0 & 0 & 0 & 0 & 0 & 0  & 0 & \frac{8}{3}\nu & 0   \\
0  & \frac{2}{3}\nu & -2\nu  & -2\nu  & 0 & 0 & 0 & 0 & 0 & 0 & 0  & 0 & 0 & \frac{8}{3}\nu   \\
0  & -2\nu & \frac{2}{3}\nu  & \frac{2}{3}\nu  & 0 & 0 & 0 & 0 & 0 & 0 & 0  & 0 & 0 & 0   \\
-2\nu  & 0 & \frac{2}{3}\nu  & \frac{2}{3}\nu  & 0 & 0 & 0 & 0 & 0 & 0 & 0  & 0 & 0 & 0   \\
\end{array}
\right),
\end{equation*}
\end{widetext}
\begin{widetext}
\begin{equation*}
\hat{S}_1=
\left(
\begin{array}{cccccccccccccc}
0 & 0  & 0  & 0  & -\frac{1}{3}\theta & 0 &  1  &  C^0_q  &  \frac{1}{3}S^0_q  &  0  &  0  &  -S^0_q  &  -\frac{2}{3}\phi  &  0    \\
   & 0 & 0  & 0  &  0 & -\frac{1}{3}\theta &  C^0_q   & 1  &  0 &  \frac{1}{3}S^0_q &  -S^0_q & 0  &  0 & -\frac{2}{3}\phi   \\
   &  & -\frac{4}{3}\gamma_\qq & -4\gamma_\qq  & \theta & \theta &  -\frac{1}{3}\theta &  -\frac{1}{3}\theta & -S^0_q  & -S^0_q
     & \frac{1}{3}S^0_q  & \frac{1}{3}S^0_q  &  2\phi &  2\phi   \\
   &  &  & -\frac{4}{3}\gamma_\qq & \theta  & \theta   &  -\frac{1}{3}\theta &  -\frac{1}{3}\theta & -S^0_q  &  -S^0_q
    & \frac{1}{3}S^0_q  &  \frac{1}{3}S^0_q &  2\phi & 2\phi    \\
  &  &   &   & -\frac{1}{3}\theta & 0 & 0 & 0 & \frac{4}{3}S^0_q & 0 & 0  & 0 & 0 & 0   \\
  &  &  &  &  & -\frac{1}{3}\theta  & 0 &  0 &  0 & \frac{4}{3}S^0_q  &  0 & 0 & 0 & 0   \\
   &   &   &   &   &   & -\frac{4}{3} &  0 & 0  &  0 & 0  & \frac{4}{3}S^0_q  & 0  & 0  \\
   &   &   &   &   &   &   & -\frac{4}{3}  & 0  & 0  & -\frac{4}{3}  & 0  &  0 & 0    \\
   &   &   &   &   &   &   &   & -\frac{1}{3}\vartheta  & 0  & 0  & 0  & 0  &  0   \\
   &   &   &   &   &   &   &   &   & -\frac{1}{3}\vartheta  & 0  &  0 & 0  &  0    \\
   &   &   &   &   &   &   &   &   &   & -\frac{4}{3} & \frac{4}{3}C^0_q  & 0  & 0    \\
   &   &   &   &   &   &   &   &   &   &   & -\frac{4}{3} &  0 &  0   \\
   &   &   &   &   &   &   &   &   &   &   &   & -\frac{8}{3}\phi & 0    \\
   &   &   &   &   &   &   &   &   &   &   &   &   & -\frac{8}{3}\phi   \\
\end{array}
\right),
\end{equation*}
\end{widetext}
\begin{widetext}
\begin{equation*}
\hat{S}_2=
\left(
\begin{array}{cccccccccccccc}
-\frac{8}{3} & -\frac{8}{3}\lambda  & 0  & 0  & 0 & -\frac{8}{3}\chi &  0  &  0  &  0  &  \frac{8}{3}\lambda  &  0  &  0  &  0  &  \frac{8}{3}\nu    \\
   & -\frac{8}{3} & 0  & 0  &  -\frac{8}{3}\chi & 0 &  0   & 0  &  \frac{8}{3}\mu &  0 &  0 & 0  &  \frac{8}{3}\nu & 0   \\
   &  & -\frac{8}{3}\theta & 0  & 0 &  0 &  -\frac{8}{3}\nu & 0  & 0 & 0 & -\frac{8}{3}\mu  &  0 & 0  & 0 \\
   &  &  & -\frac{8}{3}\theta & 0  & 0   & 0 & -\frac{8}{3}\nu &0  & 0 & 0  & -\frac{8}{3}\mu &  0 & 0    \\
  &  &   &   & -\frac{8}{3} & -\frac{8}{3}\lambda & 0 & 0 & 0 & -\frac{8}{3}\nu & 0  & -\frac{8}{3}\mu & 0 & 0   \\
  &  &  &  &  & -\frac{8}{3}  & 0 &  0 &  -\frac{8}{3}\nu & 0  &  0 & 0 & -\frac{8}{3}\mu & 0   \\
   &   &   &   &   &   & \frac{8}{3}\varphi &  0 & 0  &  0 & -\frac{8}{3}\chi  & 0  & 0  & 0  \\
   &   &   &   &   &   &   & \frac{8}{3}\varphi  & 0  & 0  & 0  & -\frac{8}{3}\chi  &  0 & 0    \\
   &   &   &   &   &   &   &   & -\frac{8}{3}  & -\frac{8}{3}\lambda  & 0  & 0  & 0  &  -\frac{8}{3}\chi  \\
   &   &   &   &   &   &   &   &   & -\frac{8}{3}  & 0  &  0 & -\frac{8}{3}\chi  &  0    \\
   &   &   &   &   &   &   &   &   &   & \frac{8}{3}\varphi & 0  & 0  & 0    \\
   &   &   &   &   &   &   &   &   &   &   & \frac{8}{3}\varphi &  0 &  0   \\
   &   &   &   &   &   &   &   &   &   &   &   & -\frac{8}{3} & \frac{8}{3}\lambda    \\
   &   &   &   &   &   &   &   &   &   &   &   &   & -\frac{8}{3}   \\
\end{array}
\right).
\end{equation*}
\end{widetext}
In the above we have introduced the following notations:
\begin{eqnarray}
 C^0_q&=&\cos q_x,\ C^1_q=\cos \frac{q_x}{2},\ C^2_q=\cos \frac{\sqrt{3}}{2}q_y,\\
 S^0_q&=&\cos q_x,\ S^1_q=\cos \frac{q_x}{2},\ S^2_q=\cos \frac{\sqrt{3}}{2}q_y,\\
 \lambda&=&C^1_qC^2_q,\ \mu=C^1_qS^2_q,\ \nu=S^1_qC^2_q,\ \chi=S^1_qS^2_q,\\
 \theta&=&C^0_q+1,\ \vartheta=C^0_q-1,\\
 \phi&=&C^1_qC^2_q+1,\ \varphi=C^1_qC^2_q-1.
\end{eqnarray}
Only the upper right parts of $\hat{S}_1$ and $\hat{S}_2$ are shown since the matrices are symmetrical.

In Fig.~\ref{fig:rixsv4} we show the results of our computation for $S=1/2$ and $\alpha=1$. The numerical solution of Eq.~(\ref{Eq:fincorr}) by considering only the $\mathcal{V}_4$ vertex
is performed with $69\times69$ lattice sites, while the integrals arising in the exact solution are solved on a mesh of size $252\times252$ and then extrapolated to $N\rightarrow\infty$ ~\cite{PhysRevB.89.165103}.
\bibliography{reftriangular,trimaterials,triangleextraref}

\begin{thebibliography}{66}%
\makeatletter
\providecommand \@ifxundefined [1]{%
 \@ifx{#1\undefined}
}%
\providecommand \@ifnum [1]{%
 \ifnum #1\expandafter \@firstoftwo
 \else \expandafter \@secondoftwo
 \fi
}%
\providecommand \@ifx [1]{%
 \ifx #1\expandafter \@firstoftwo
 \else \expandafter \@secondoftwo
 \fi
}%
\providecommand \natexlab [1]{#1}%
\providecommand \enquote  [1]{``#1''}%
\providecommand \bibnamefont  [1]{#1}%
\providecommand \bibfnamefont [1]{#1}%
\providecommand \citenamefont [1]{#1}%
\providecommand \href@noop [0]{\@secondoftwo}%
\providecommand \href [0]{\begingroup \@sanitize@url \@href}%
\providecommand \@href[1]{\@@startlink{#1}\@@href}%
\providecommand \@@href[1]{\endgroup#1\@@endlink}%
\providecommand \@sanitize@url [0]{\catcode `\\12\catcode `\$12\catcode
  `\&12\catcode `\#12\catcode `\^12\catcode `\_12\catcode `\%12\relax}%
\providecommand \@@startlink[1]{}%
\providecommand \@@endlink[0]{}%
\providecommand \url  [0]{\begingroup\@sanitize@url \@url }%
\providecommand \@url [1]{\endgroup\@href {#1}{\urlprefix }}%
\providecommand \urlprefix  [0]{URL }%
\providecommand \Eprint [0]{\href }%
\providecommand \doibase [0]{http://dx.doi.org/}%
\providecommand \selectlanguage [0]{\@gobble}%
\providecommand \bibinfo  [0]{\@secondoftwo}%
\providecommand \bibfield  [0]{\@secondoftwo}%
\providecommand \translation [1]{[#1]}%
\providecommand \BibitemOpen [0]{}%
\providecommand \bibitemStop [0]{}%
\providecommand \bibitemNoStop [0]{.\EOS\space}%
\providecommand \EOS [0]{\spacefactor3000\relax}%
\providecommand \BibitemShut  [1]{\csname bibitem#1\endcsname}%
\let\auto@bib@innerbib\@empty
\bibitem [{\citenamefont {Anderson}(1973)}]{Anderson1973153}%
  \BibitemOpen
  \bibfield  {author} {\bibinfo {author} {\bibfnamefont {P.~W.}\ \bibnamefont
  {Anderson}},\ }\bibfield  {title} {\enquote {\bibinfo {title} {Resonating
  valence bonds: A new kind of insulator?}}\ }\href {\doibase
  10.1016/0025-5408(73)90167-0} {\bibfield  {journal} {\bibinfo  {journal}
  {Mater. Res. Bull.}\ }\textbf {\bibinfo {volume} {8}},\ \bibinfo {pages}
  {153} (\bibinfo {year} {1973})}\BibitemShut {NoStop}%
\bibitem [{\citenamefont {Jolicoeur}\ and\ \citenamefont
  {Le~Guillou}(1989)}]{PhysRevB.40.2727}%
  \BibitemOpen
  \bibfield  {author} {\bibinfo {author} {\bibfnamefont {Th.}\ \bibnamefont
  {Jolicoeur}}\ and\ \bibinfo {author} {\bibfnamefont {J.~C.}\ \bibnamefont
  {Le~Guillou}},\ }\bibfield  {title} {\enquote {\bibinfo {title} {Spin-wave
  results for the triangular heisenberg antiferromagnet},}\ }\href {\doibase
  10.1103/PhysRevB.40.2727} {\bibfield  {journal} {\bibinfo  {journal} {Phys.
  Rev. B}\ }\textbf {\bibinfo {volume} {40}},\ \bibinfo {pages} {2727--2729}
  (\bibinfo {year} {1989})}\BibitemShut {NoStop}%
\bibitem [{\citenamefont {Miyake}(1992)}]{JPSJ.61.983}%
  \BibitemOpen
  \bibfield  {author} {\bibinfo {author} {\bibfnamefont {Satoru~J.}\
  \bibnamefont {Miyake}},\ }\bibfield  {title} {\enquote {\bibinfo {title}
  {Spin-wave results for the staggered magnetization of triangular heisenberg
  antiferromagnet},}\ }\href {\doibase 10.1143/JPSJ.61.983} {\bibfield
  {journal} {\bibinfo  {journal} {J. Phys. Soc. Jpn.}\ }\textbf {\bibinfo
  {volume} {61}},\ \bibinfo {pages} {983--988} (\bibinfo {year}
  {1992})}\BibitemShut {NoStop}%
\bibitem [{\citenamefont {Chubukov}\ \emph {et~al.}(1994)\citenamefont
  {Chubukov}, \citenamefont {Sachdev},\ and\ \citenamefont
  {Senthil}}]{JPCM.6.8891}%
  \BibitemOpen
  \bibfield  {author} {\bibinfo {author} {\bibfnamefont {A~V}\ \bibnamefont
  {Chubukov}}, \bibinfo {author} {\bibfnamefont {S}~\bibnamefont {Sachdev}}, \
  and\ \bibinfo {author} {\bibfnamefont {T}~\bibnamefont {Senthil}},\
  }\bibfield  {title} {\enquote {\bibinfo {title} {Large-s expansion for
  quantum antiferromagnets on a triangular lattice},}\ }\href
  {http://stacks.iop.org/0953-8984/6/i=42/a=019} {\bibfield  {journal}
  {\bibinfo  {journal} {J. Phys.: Condens. Matter}\ }\textbf {\bibinfo {volume}
  {6}},\ \bibinfo {pages} {8891} (\bibinfo {year} {1994})}\BibitemShut
  {NoStop}%
\bibitem [{\citenamefont {Capriotti}\ \emph {et~al.}(1999)\citenamefont
  {Capriotti}, \citenamefont {Trumper},\ and\ \citenamefont
  {Sorella}}]{PhysRevLett.82.3899}%
  \BibitemOpen
  \bibfield  {author} {\bibinfo {author} {\bibfnamefont {Luca}\ \bibnamefont
  {Capriotti}}, \bibinfo {author} {\bibfnamefont {Adolfo~E.}\ \bibnamefont
  {Trumper}}, \ and\ \bibinfo {author} {\bibfnamefont {Sandro}\ \bibnamefont
  {Sorella}},\ }\bibfield  {title} {\enquote {\bibinfo {title} {Long-range
  n\'eel order in the triangular heisenberg model},}\ }\href {\doibase
  10.1103/PhysRevLett.82.3899} {\bibfield  {journal} {\bibinfo  {journal}
  {Phys. Rev. Lett.}\ }\textbf {\bibinfo {volume} {82}},\ \bibinfo {pages}
  {3899--3902} (\bibinfo {year} {1999})}\BibitemShut {NoStop}%
\bibitem [{\citenamefont {Zheng}\ \emph
  {et~al.}(2006{\natexlab{a}})\citenamefont {Zheng}, \citenamefont
  {Fj\ae{}restad}, \citenamefont {Singh}, \citenamefont {McKenzie},\ and\
  \citenamefont {Coldea}}]{PhysRevB.74.224420}%
  \BibitemOpen
  \bibfield  {author} {\bibinfo {author} {\bibfnamefont {Weihong}\ \bibnamefont
  {Zheng}}, \bibinfo {author} {\bibfnamefont {John~O.}\ \bibnamefont
  {Fj\ae{}restad}}, \bibinfo {author} {\bibfnamefont {Rajiv R.~P.}\
  \bibnamefont {Singh}}, \bibinfo {author} {\bibfnamefont {Ross~H.}\
  \bibnamefont {McKenzie}}, \ and\ \bibinfo {author} {\bibfnamefont {Radu}\
  \bibnamefont {Coldea}},\ }\bibfield  {title} {\enquote {\bibinfo {title}
  {Excitation spectra of the spin-$\frac{1}{2}$ triangular-lattice heisenberg
  antiferromagnet},}\ }\href {\doibase 10.1103/PhysRevB.74.224420} {\bibfield
  {journal} {\bibinfo  {journal} {Phys. Rev. B}\ }\textbf {\bibinfo {volume}
  {74}},\ \bibinfo {pages} {224420} (\bibinfo {year}
  {2006}{\natexlab{a}})}\BibitemShut {NoStop}%
\bibitem [{\citenamefont {White}\ and\ \citenamefont
  {Chernyshev}(2007)}]{PhysRevLett.99.127004}%
  \BibitemOpen
  \bibfield  {author} {\bibinfo {author} {\bibfnamefont {Steven~R.}\
  \bibnamefont {White}}\ and\ \bibinfo {author} {\bibfnamefont {A.~L.}\
  \bibnamefont {Chernyshev}},\ }\bibfield  {title} {\enquote {\bibinfo {title}
  {Ne\'el order in square and triangular lattice heisenberg models},}\ }\href
  {\doibase 10.1103/PhysRevLett.99.127004} {\bibfield  {journal} {\bibinfo
  {journal} {Phys. Rev. Lett.}\ }\textbf {\bibinfo {volume} {99}},\ \bibinfo
  {pages} {127004} (\bibinfo {year} {2007})}\BibitemShut {NoStop}%
\bibitem [{\citenamefont {Li}\ \emph {et~al.}(2015)\citenamefont {Li},
  \citenamefont {Bishop},\ and\ \citenamefont
  {Campbell}}]{LiPhysRevB.91.014426}%
  \BibitemOpen
  \bibfield  {author} {\bibinfo {author} {\bibfnamefont {P.~H.~Y.}\
  \bibnamefont {Li}}, \bibinfo {author} {\bibfnamefont {R.~F.}\ \bibnamefont
  {Bishop}}, \ and\ \bibinfo {author} {\bibfnamefont {C.~E.}\ \bibnamefont
  {Campbell}},\ }\bibfield  {title} {\enquote {\bibinfo {title} {Quasiclassical
  magnetic order and its loss in a spin-$\frac{1}{2}$ heisenberg
  antiferromagnet on a triangular lattice with competing bonds},}\ }\href
  {\doibase 10.1103/PhysRevB.91.014426} {\bibfield  {journal} {\bibinfo
  {journal} {Phys. Rev. B}\ }\textbf {\bibinfo {volume} {91}},\ \bibinfo
  {pages} {014426} (\bibinfo {year} {2015})}\BibitemShut {NoStop}%
\bibitem [{\citenamefont {Huse}\ and\ \citenamefont
  {Elser}(1988)}]{HusePhysRevLett.60.2531}%
  \BibitemOpen
  \bibfield  {author} {\bibinfo {author} {\bibfnamefont {David~A.}\
  \bibnamefont {Huse}}\ and\ \bibinfo {author} {\bibfnamefont {Veit}\
  \bibnamefont {Elser}},\ }\bibfield  {title} {\enquote {\bibinfo {title}
  {Simple variational wave functions for two-dimensional heisenberg
  spin-\textonehalf{} antiferromagnets},}\ }\href {\doibase
  10.1103/PhysRevLett.60.2531} {\bibfield  {journal} {\bibinfo  {journal}
  {Phys. Rev. Lett.}\ }\textbf {\bibinfo {volume} {60}},\ \bibinfo {pages}
  {2531--2534} (\bibinfo {year} {1988})}\BibitemShut {NoStop}%
\bibitem [{\citenamefont {Bernu}\ \emph {et~al.}(1994)\citenamefont {Bernu},
  \citenamefont {Lecheminant}, \citenamefont {Lhuillier},\ and\ \citenamefont
  {Pierre}}]{BernuPhysRevB.50.10048}%
  \BibitemOpen
  \bibfield  {author} {\bibinfo {author} {\bibfnamefont {B.}~\bibnamefont
  {Bernu}}, \bibinfo {author} {\bibfnamefont {P.}~\bibnamefont {Lecheminant}},
  \bibinfo {author} {\bibfnamefont {C.}~\bibnamefont {Lhuillier}}, \ and\
  \bibinfo {author} {\bibfnamefont {L.}~\bibnamefont {Pierre}},\ }\bibfield
  {title} {\enquote {\bibinfo {title} {Exact spectra, spin susceptibilities,
  and order parameter of the quantum heisenberg antiferromagnet on the
  triangular lattice},}\ }\href {\doibase 10.1103/PhysRevB.50.10048} {\bibfield
   {journal} {\bibinfo  {journal} {Phys. Rev. B}\ }\textbf {\bibinfo {volume}
  {50}},\ \bibinfo {pages} {10048--10062} (\bibinfo {year} {1994})}\BibitemShut
  {NoStop}%
\bibitem [{\citenamefont {Singh}\ and\ \citenamefont
  {Huse}(1992)}]{SinghPhysRevLett.68.1766}%
  \BibitemOpen
  \bibfield  {author} {\bibinfo {author} {\bibfnamefont {Rajiv R.~P.}\
  \bibnamefont {Singh}}\ and\ \bibinfo {author} {\bibfnamefont {David~A.}\
  \bibnamefont {Huse}},\ }\bibfield  {title} {\enquote {\bibinfo {title}
  {Three-sublattice order in triangular- and kagom\'e-lattice spin-half
  antiferromagnets},}\ }\href {\doibase 10.1103/PhysRevLett.68.1766} {\bibfield
   {journal} {\bibinfo  {journal} {Phys. Rev. Lett.}\ }\textbf {\bibinfo
  {volume} {68}},\ \bibinfo {pages} {1766--1769} (\bibinfo {year}
  {1992})}\BibitemShut {NoStop}%
\bibitem [{\citenamefont {Shirata}\ \emph {et~al.}(2012)\citenamefont
  {Shirata}, \citenamefont {Tanaka}, \citenamefont {Matsuo},\ and\
  \citenamefont {Kindo}}]{PhysRevLett.108.057205}%
  \BibitemOpen
  \bibfield  {author} {\bibinfo {author} {\bibfnamefont {Yutaka}\ \bibnamefont
  {Shirata}}, \bibinfo {author} {\bibfnamefont {Hidekazu}\ \bibnamefont
  {Tanaka}}, \bibinfo {author} {\bibfnamefont {Akira}\ \bibnamefont {Matsuo}},
  \ and\ \bibinfo {author} {\bibfnamefont {Koichi}\ \bibnamefont {Kindo}},\
  }\bibfield  {title} {\enquote {\bibinfo {title} {Experimental realization of
  a spin-$1/2$ triangular-lattice heisenberg antiferromagnet},}\ }\href
  {\doibase 10.1103/PhysRevLett.108.057205} {\bibfield  {journal} {\bibinfo
  {journal} {Phys. Rev. Lett.}\ }\textbf {\bibinfo {volume} {108}},\ \bibinfo
  {pages} {057205} (\bibinfo {year} {2012})}\BibitemShut {NoStop}%
\bibitem [{\citenamefont {Koutroulakis}\ \emph {et~al.}(2015)\citenamefont
  {Koutroulakis}, \citenamefont {Zhou}, \citenamefont {Kamiya}, \citenamefont
  {Thompson}, \citenamefont {Zhou}, \citenamefont {Batista},\ and\
  \citenamefont {Brown}}]{PhysRevB.91.024410}%
  \BibitemOpen
  \bibfield  {author} {\bibinfo {author} {\bibfnamefont {G.}~\bibnamefont
  {Koutroulakis}}, \bibinfo {author} {\bibfnamefont {T.}~\bibnamefont {Zhou}},
  \bibinfo {author} {\bibfnamefont {Y.}~\bibnamefont {Kamiya}}, \bibinfo
  {author} {\bibfnamefont {J.~D.}\ \bibnamefont {Thompson}}, \bibinfo {author}
  {\bibfnamefont {H.~D.}\ \bibnamefont {Zhou}}, \bibinfo {author}
  {\bibfnamefont {C.~D.}\ \bibnamefont {Batista}}, \ and\ \bibinfo {author}
  {\bibfnamefont {S.~E.}\ \bibnamefont {Brown}},\ }\bibfield  {title} {\enquote
  {\bibinfo {title} {Quantum phase diagram of the $s=\frac{1}{2}$
  triangular-lattice antiferromagnet
  ${\mathrm{ba}}_{3}{\mathrm{cosb}}_{2}{\mathrm{o}}_{9}$},}\ }\href {\doibase
  10.1103/PhysRevB.91.024410} {\bibfield  {journal} {\bibinfo  {journal} {Phys.
  Rev. B}\ }\textbf {\bibinfo {volume} {91}},\ \bibinfo {pages} {024410}
  (\bibinfo {year} {2015})}\BibitemShut {NoStop}%
\bibitem [{\citenamefont {Susuki}\ \emph {et~al.}(2013)\citenamefont {Susuki},
  \citenamefont {Kurita}, \citenamefont {Tanaka}, \citenamefont {Nojiri},
  \citenamefont {Matsuo}, \citenamefont {Kindo},\ and\ \citenamefont
  {Tanaka}}]{SusukiPhysRevLett.110.267201}%
  \BibitemOpen
  \bibfield  {author} {\bibinfo {author} {\bibfnamefont {Takuya}\ \bibnamefont
  {Susuki}}, \bibinfo {author} {\bibfnamefont {Nobuyuki}\ \bibnamefont
  {Kurita}}, \bibinfo {author} {\bibfnamefont {Takuya}\ \bibnamefont {Tanaka}},
  \bibinfo {author} {\bibfnamefont {Hiroyuki}\ \bibnamefont {Nojiri}}, \bibinfo
  {author} {\bibfnamefont {Akira}\ \bibnamefont {Matsuo}}, \bibinfo {author}
  {\bibfnamefont {Koichi}\ \bibnamefont {Kindo}}, \ and\ \bibinfo {author}
  {\bibfnamefont {Hidekazu}\ \bibnamefont {Tanaka}},\ }\bibfield  {title}
  {\enquote {\bibinfo {title} {Magnetization process and collective excitations
  in the $s\mathbf{=}1/2$ triangular-lattice heisenberg antiferromagnet
  ${\mathrm{ba}}_{3}{\mathrm{cosb}}_{2}{\mathbf{o}}_{9}$},}\ }\href {\doibase
  10.1103/PhysRevLett.110.267201} {\bibfield  {journal} {\bibinfo  {journal}
  {Phys. Rev. Lett.}\ }\textbf {\bibinfo {volume} {110}},\ \bibinfo {pages}
  {267201} (\bibinfo {year} {2013})}\BibitemShut {NoStop}%
\bibitem [{\citenamefont {Ono}\ \emph {et~al.}(2003)\citenamefont {Ono},
  \citenamefont {Tanaka}, \citenamefont {Aruga~Katori}, \citenamefont
  {Ishikawa}, \citenamefont {Mitamura},\ and\ \citenamefont
  {Goto}}]{OnoPhysRevB.67.104431}%
  \BibitemOpen
  \bibfield  {author} {\bibinfo {author} {\bibfnamefont {T.}~\bibnamefont
  {Ono}}, \bibinfo {author} {\bibfnamefont {H.}~\bibnamefont {Tanaka}},
  \bibinfo {author} {\bibfnamefont {H.}~\bibnamefont {Aruga~Katori}}, \bibinfo
  {author} {\bibfnamefont {F.}~\bibnamefont {Ishikawa}}, \bibinfo {author}
  {\bibfnamefont {H.}~\bibnamefont {Mitamura}}, \ and\ \bibinfo {author}
  {\bibfnamefont {T.}~\bibnamefont {Goto}},\ }\bibfield  {title} {\enquote
  {\bibinfo {title} {Magnetization plateau in the frustrated quantum spin
  system ${\mathrm{cs}}_{2}{\mathrm{cubr}}_{4}$},}\ }\href {\doibase
  10.1103/PhysRevB.67.104431} {\bibfield  {journal} {\bibinfo  {journal} {Phys.
  Rev. B}\ }\textbf {\bibinfo {volume} {67}},\ \bibinfo {pages} {104431}
  (\bibinfo {year} {2003})}\BibitemShut {NoStop}%
\bibitem [{\citenamefont {Kadowaki}\ \emph {et~al.}(1987)\citenamefont
  {Kadowaki}, \citenamefont {Ubukoshi}, \citenamefont {Hirakawa}, \citenamefont
  {Martínez},\ and\ \citenamefont {Shirane}}]{KadowakiJPSJ}%
  \BibitemOpen
  \bibfield  {author} {\bibinfo {author} {\bibfnamefont {Hiroaki}\ \bibnamefont
  {Kadowaki}}, \bibinfo {author} {\bibfnamefont {Koji}\ \bibnamefont
  {Ubukoshi}}, \bibinfo {author} {\bibfnamefont {Kinshiro}\ \bibnamefont
  {Hirakawa}}, \bibinfo {author} {\bibfnamefont {José~L.}\ \bibnamefont
  {Martínez}}, \ and\ \bibinfo {author} {\bibfnamefont {Gen}\ \bibnamefont
  {Shirane}},\ }\bibfield  {title} {\enquote {\bibinfo {title} {Experimental
  study of new type phase transition in triangular lattice antiferromagnet
  vcl2},}\ }\href {\doibase 10.1143/JPSJ.56.4027} {\bibfield  {journal}
  {\bibinfo  {journal} {J. Phys. Soc. Jpn}\ }\textbf {\bibinfo {volume} {56}},\
  \bibinfo {pages} {4027--4039} (\bibinfo {year} {1987})}\BibitemShut {NoStop}%
\bibitem [{\citenamefont {Ishii}\ \emph {et~al.}(2011)\citenamefont {Ishii},
  \citenamefont {Tanaka}, \citenamefont {Onuma}, \citenamefont {Nambu},
  \citenamefont {Tokunaga}, \citenamefont {Sakakibara}, \citenamefont
  {Kawashima}, \citenamefont {Maeno}, \citenamefont {Broholm}, \citenamefont
  {Gautreaux}, \citenamefont {Chan},\ and\ \citenamefont
  {Nakatsuji}}]{IshiiEPL}%
  \BibitemOpen
  \bibfield  {author} {\bibinfo {author} {\bibfnamefont {R.}~\bibnamefont
  {Ishii}}, \bibinfo {author} {\bibfnamefont {S.}~\bibnamefont {Tanaka}},
  \bibinfo {author} {\bibfnamefont {K.}~\bibnamefont {Onuma}}, \bibinfo
  {author} {\bibfnamefont {Y.}~\bibnamefont {Nambu}}, \bibinfo {author}
  {\bibfnamefont {M.}~\bibnamefont {Tokunaga}}, \bibinfo {author}
  {\bibfnamefont {T.}~\bibnamefont {Sakakibara}}, \bibinfo {author}
  {\bibfnamefont {N.}~\bibnamefont {Kawashima}}, \bibinfo {author}
  {\bibfnamefont {Y.}~\bibnamefont {Maeno}}, \bibinfo {author} {\bibfnamefont
  {C.}~\bibnamefont {Broholm}}, \bibinfo {author} {\bibfnamefont {D.~P.}\
  \bibnamefont {Gautreaux}}, \bibinfo {author} {\bibfnamefont {J.~Y.}\
  \bibnamefont {Chan}}, \ and\ \bibinfo {author} {\bibfnamefont
  {S.}~\bibnamefont {Nakatsuji}},\ }\bibfield  {title} {\enquote {\bibinfo
  {title} {Successive phase transitions and phase diagrams for the
  quasi-two-dimensional easy-axis triangular antiferromagnet rb 4 mn(moo 4 )
  3},}\ }\href {http://stacks.iop.org/0295-5075/94/i=1/a=17001} {\bibfield
  {journal} {\bibinfo  {journal} {Eur. Phys. Lett.}\ }\textbf {\bibinfo
  {volume} {94}},\ \bibinfo {pages} {17001} (\bibinfo {year}
  {2011})}\BibitemShut {NoStop}%
\bibitem [{\citenamefont {Poienar}\ \emph {et~al.}(2010)\citenamefont
  {Poienar}, \citenamefont {Damay}, \citenamefont {Martin}, \citenamefont
  {Robert},\ and\ \citenamefont {Petit}}]{PetitPhysRevB.81.104411}%
  \BibitemOpen
  \bibfield  {author} {\bibinfo {author} {\bibfnamefont {M.}~\bibnamefont
  {Poienar}}, \bibinfo {author} {\bibfnamefont {F.}~\bibnamefont {Damay}},
  \bibinfo {author} {\bibfnamefont {C.}~\bibnamefont {Martin}}, \bibinfo
  {author} {\bibfnamefont {J.}~\bibnamefont {Robert}}, \ and\ \bibinfo {author}
  {\bibfnamefont {S.}~\bibnamefont {Petit}},\ }\bibfield  {title} {\enquote
  {\bibinfo {title} {Spin dynamics in the geometrically frustrated multiferroic
  ${\text{cucro}}_{2}$},}\ }\href {\doibase 10.1103/PhysRevB.81.104411}
  {\bibfield  {journal} {\bibinfo  {journal} {Phys. Rev. B}\ }\textbf {\bibinfo
  {volume} {81}},\ \bibinfo {pages} {104411} (\bibinfo {year}
  {2010})}\BibitemShut {NoStop}%
\bibitem [{\citenamefont {Toth}\ \emph {et~al.}(2011)\citenamefont {Toth},
  \citenamefont {Lake}, \citenamefont {Kimber}, \citenamefont {Pieper},
  \citenamefont {Reehuis}, \citenamefont {Islam}, \citenamefont {Zaharko},
  \citenamefont {Ritter}, \citenamefont {Hill}, \citenamefont {Ryll},
  \citenamefont {Kiefer}, \citenamefont {Argyriou},\ and\ \citenamefont
  {Williams}}]{TothPhysRevB.84.054452}%
  \BibitemOpen
  \bibfield  {author} {\bibinfo {author} {\bibfnamefont {S.}~\bibnamefont
  {Toth}}, \bibinfo {author} {\bibfnamefont {B.}~\bibnamefont {Lake}}, \bibinfo
  {author} {\bibfnamefont {S.~A.~J.}\ \bibnamefont {Kimber}}, \bibinfo {author}
  {\bibfnamefont {O.}~\bibnamefont {Pieper}}, \bibinfo {author} {\bibfnamefont
  {M.}~\bibnamefont {Reehuis}}, \bibinfo {author} {\bibfnamefont {A.~T. M.~N.}\
  \bibnamefont {Islam}}, \bibinfo {author} {\bibfnamefont {O.}~\bibnamefont
  {Zaharko}}, \bibinfo {author} {\bibfnamefont {C.}~\bibnamefont {Ritter}},
  \bibinfo {author} {\bibfnamefont {A.~H.}\ \bibnamefont {Hill}}, \bibinfo
  {author} {\bibfnamefont {H.}~\bibnamefont {Ryll}}, \bibinfo {author}
  {\bibfnamefont {K.}~\bibnamefont {Kiefer}}, \bibinfo {author} {\bibfnamefont
  {D.~N.}\ \bibnamefont {Argyriou}}, \ and\ \bibinfo {author} {\bibfnamefont
  {A.~J.}\ \bibnamefont {Williams}},\ }\bibfield  {title} {\enquote {\bibinfo
  {title} {120${}^{\circ}$ helical magnetic order in the distorted triangular
  antiferromagnet $\alpha\text{-}{\mathrm{ca}}_{2}{\mathrm{cubr}}_{4}$},}\
  }\href {\doibase 10.1103/PhysRevB.84.054452} {\bibfield  {journal} {\bibinfo
  {journal} {Phys. Rev. B}\ }\textbf {\bibinfo {volume} {84}},\ \bibinfo
  {pages} {054452} (\bibinfo {year} {2011})}\BibitemShut {NoStop}%
\bibitem [{\citenamefont {Toth}\ \emph {et~al.}(2012)\citenamefont {Toth},
  \citenamefont {Lake}, \citenamefont {Hradil}, \citenamefont {Guidi},
  \citenamefont {Rule}, \citenamefont {Stone},\ and\ \citenamefont
  {Islam}}]{TothPhysRevLett.109.127203}%
  \BibitemOpen
  \bibfield  {author} {\bibinfo {author} {\bibfnamefont {S.}~\bibnamefont
  {Toth}}, \bibinfo {author} {\bibfnamefont {B.}~\bibnamefont {Lake}}, \bibinfo
  {author} {\bibfnamefont {K.}~\bibnamefont {Hradil}}, \bibinfo {author}
  {\bibfnamefont {T.}~\bibnamefont {Guidi}}, \bibinfo {author} {\bibfnamefont
  {K.~C.}\ \bibnamefont {Rule}}, \bibinfo {author} {\bibfnamefont {M.~B.}\
  \bibnamefont {Stone}}, \ and\ \bibinfo {author} {\bibfnamefont {A.~T. M.~N.}\
  \bibnamefont {Islam}},\ }\bibfield  {title} {\enquote {\bibinfo {title}
  {Magnetic soft modes in the distorted triangular antiferromagnet
  $\alpha\text{-}{\mathrm{ca}}_{2}{\mathrm{cubr}}_{4}$},}\ }\href {\doibase
  10.1103/PhysRevLett.109.127203} {\bibfield  {journal} {\bibinfo  {journal}
  {Phys. Rev. Lett.}\ }\textbf {\bibinfo {volume} {109}},\ \bibinfo {pages}
  {127203} (\bibinfo {year} {2012})}\BibitemShut {NoStop}%
\bibitem [{\citenamefont {Coldea}\ \emph
  {et~al.}(2001{\natexlab{a}})\citenamefont {Coldea}, \citenamefont {Tennant},
  \citenamefont {Tsvelik},\ and\ \citenamefont
  {Tylczynski}}]{PhysRevLett.86.1335}%
  \BibitemOpen
  \bibfield  {author} {\bibinfo {author} {\bibfnamefont {R.}~\bibnamefont
  {Coldea}}, \bibinfo {author} {\bibfnamefont {D.~A.}\ \bibnamefont {Tennant}},
  \bibinfo {author} {\bibfnamefont {A.~M.}\ \bibnamefont {Tsvelik}}, \ and\
  \bibinfo {author} {\bibfnamefont {Z.}~\bibnamefont {Tylczynski}},\ }\bibfield
   {title} {\enquote {\bibinfo {title} {Experimental realization of a 2d
  fractional quantum spin liquid},}\ }\href {\doibase
  10.1103/PhysRevLett.86.1335} {\bibfield  {journal} {\bibinfo  {journal}
  {Phys. Rev. Lett.}\ }\textbf {\bibinfo {volume} {86}},\ \bibinfo {pages}
  {1335--1338} (\bibinfo {year} {2001}{\natexlab{a}})}\BibitemShut {NoStop}%
\bibitem [{\citenamefont {Chen}\ \emph {et~al.}(2013)\citenamefont {Chen},
  \citenamefont {Ju}, \citenamefont {Jiang}, \citenamefont {Starykh},\ and\
  \citenamefont {Balents}}]{ChenPhysRevB.87.165123}%
  \BibitemOpen
  \bibfield  {author} {\bibinfo {author} {\bibfnamefont {Ru}~\bibnamefont
  {Chen}}, \bibinfo {author} {\bibfnamefont {Hyejin}\ \bibnamefont {Ju}},
  \bibinfo {author} {\bibfnamefont {Hong-Chen}\ \bibnamefont {Jiang}}, \bibinfo
  {author} {\bibfnamefont {Oleg~A.}\ \bibnamefont {Starykh}}, \ and\ \bibinfo
  {author} {\bibfnamefont {Leon}\ \bibnamefont {Balents}},\ }\bibfield  {title}
  {\enquote {\bibinfo {title} {Ground states of spin-$\frac{1}{2}$ triangular
  antiferromagnets in a magnetic field},}\ }\href {\doibase
  10.1103/PhysRevB.87.165123} {\bibfield  {journal} {\bibinfo  {journal} {Phys.
  Rev. B}\ }\textbf {\bibinfo {volume} {87}},\ \bibinfo {pages} {165123}
  (\bibinfo {year} {2013})}\BibitemShut {NoStop}%
\bibitem [{\citenamefont {Schmidt}\ and\ \citenamefont
  {Thalmeier}(2014)}]{SchmidtPhysRevB.89.184402}%
  \BibitemOpen
  \bibfield  {author} {\bibinfo {author} {\bibfnamefont {Burkhard}\
  \bibnamefont {Schmidt}}\ and\ \bibinfo {author} {\bibfnamefont {Peter}\
  \bibnamefont {Thalmeier}},\ }\bibfield  {title} {\enquote {\bibinfo {title}
  {Quantum fluctuations in anisotropic triangular lattices with ferromagnetic
  and antiferromagnetic exchange},}\ }\href {\doibase
  10.1103/PhysRevB.89.184402} {\bibfield  {journal} {\bibinfo  {journal} {Phys.
  Rev. B}\ }\textbf {\bibinfo {volume} {89}},\ \bibinfo {pages} {184402}
  (\bibinfo {year} {2014})}\BibitemShut {NoStop}%
\bibitem [{\citenamefont {Hauke}\ \emph {et~al.}(2011)\citenamefont {Hauke},
  \citenamefont {Roscilde}, \citenamefont {Murg}, \citenamefont {Cirac},\ and\
  \citenamefont {Schmied}}]{HaukeNJP}%
  \BibitemOpen
  \bibfield  {author} {\bibinfo {author} {\bibfnamefont {Philipp}\ \bibnamefont
  {Hauke}}, \bibinfo {author} {\bibfnamefont {Tommaso}\ \bibnamefont
  {Roscilde}}, \bibinfo {author} {\bibfnamefont {Valentin}\ \bibnamefont
  {Murg}}, \bibinfo {author} {\bibfnamefont {J~Ignacio}\ \bibnamefont {Cirac}},
  \ and\ \bibinfo {author} {\bibfnamefont {Roman}\ \bibnamefont {Schmied}},\
  }\bibfield  {title} {\enquote {\bibinfo {title} {Modified spin-wave theory
  with ordering vector optimization: spatially anisotropic triangular lattice
  and j 1 j 2 j 3 model with heisenberg interactions},}\ }\href
  {http://stacks.iop.org/1367-2630/13/i=7/a=075017} {\bibfield  {journal}
  {\bibinfo  {journal} {New Journal of Physics}\ }\textbf {\bibinfo {volume}
  {13}},\ \bibinfo {pages} {075017} (\bibinfo {year} {2011})}\BibitemShut
  {NoStop}%
\bibitem [{\citenamefont {Kohno}\ \emph {et~al.}(2007)\citenamefont {Kohno},
  \citenamefont {Starykh},\ and\ \citenamefont {Balents}}]{nphys749}%
  \BibitemOpen
  \bibfield  {author} {\bibinfo {author} {\bibfnamefont {Masanori}\
  \bibnamefont {Kohno}}, \bibinfo {author} {\bibfnamefont {Oleg~A.}\
  \bibnamefont {Starykh}}, \ and\ \bibinfo {author} {\bibfnamefont {Leon}\
  \bibnamefont {Balents}},\ }\bibfield  {title} {\enquote {\bibinfo {title}
  {Spinons and triplons in spatially anisotropic frustrated
  antiferromagnets},}\ }\href {\doibase 10.1038/nphys749} {\bibfield  {journal}
  {\bibinfo  {journal} {Nature Phys.}\ }\textbf {\bibinfo {volume} {3}},\
  \bibinfo {pages} {790} (\bibinfo {year} {2007})}\BibitemShut {NoStop}%
\bibitem [{\citenamefont {Swanson}\ \emph {et~al.}(2009)\citenamefont
  {Swanson}, \citenamefont {Haraldsen},\ and\ \citenamefont
  {Fishman}}]{fishmanPhysRevB.79.184413}%
  \BibitemOpen
  \bibfield  {author} {\bibinfo {author} {\bibfnamefont {M.}~\bibnamefont
  {Swanson}}, \bibinfo {author} {\bibfnamefont {J.~T.}\ \bibnamefont
  {Haraldsen}}, \ and\ \bibinfo {author} {\bibfnamefont {R.~S.}\ \bibnamefont
  {Fishman}},\ }\bibfield  {title} {\enquote {\bibinfo {title} {Critical
  anisotropies of a geometrically frustrated triangular-lattice
  antiferromagnet},}\ }\href {\doibase 10.1103/PhysRevB.79.184413} {\bibfield
  {journal} {\bibinfo  {journal} {Phys. Rev. B}\ }\textbf {\bibinfo {volume}
  {79}},\ \bibinfo {pages} {184413} (\bibinfo {year} {2009})}\BibitemShut
  {NoStop}%
\bibitem [{\citenamefont {Fishman}\ and\ \citenamefont
  {Okamoto}(2010)}]{okamotoPhysRevB.81.020402}%
  \BibitemOpen
  \bibfield  {author} {\bibinfo {author} {\bibfnamefont {Randy~S.}\
  \bibnamefont {Fishman}}\ and\ \bibinfo {author} {\bibfnamefont {Satoshi}\
  \bibnamefont {Okamoto}},\ }\bibfield  {title} {\enquote {\bibinfo {title}
  {Noncollinear magnetic phases of a triangular-lattice antiferromagnet and of
  doped ${\mathrm{cufeo}}_{2}$},}\ }\href {\doibase 10.1103/PhysRevB.81.020402}
  {\bibfield  {journal} {\bibinfo  {journal} {Phys. Rev. B}\ }\textbf {\bibinfo
  {volume} {81}},\ \bibinfo {pages} {020402} (\bibinfo {year}
  {2010})}\BibitemShut {NoStop}%
\bibitem [{\citenamefont {Ghioldi}\ \emph {et~al.}(2015)\citenamefont
  {Ghioldi}, \citenamefont {Mezio}, \citenamefont {Manuel}, \citenamefont
  {Singh}, \citenamefont {Oitmaa},\ and\ \citenamefont
  {Trumper}}]{PhysRevB.91.134423}%
  \BibitemOpen
  \bibfield  {author} {\bibinfo {author} {\bibfnamefont {E.~A.}\ \bibnamefont
  {Ghioldi}}, \bibinfo {author} {\bibfnamefont {A.}~\bibnamefont {Mezio}},
  \bibinfo {author} {\bibfnamefont {L.~O.}\ \bibnamefont {Manuel}}, \bibinfo
  {author} {\bibfnamefont {R.~R.~P.}\ \bibnamefont {Singh}}, \bibinfo {author}
  {\bibfnamefont {J.}~\bibnamefont {Oitmaa}}, \ and\ \bibinfo {author}
  {\bibfnamefont {A.~E.}\ \bibnamefont {Trumper}},\ }\bibfield  {title}
  {\enquote {\bibinfo {title} {Magnons and excitation continuum in xxz
  triangular antiferromagnetic model: Application to
  ${\text{ba}}_{3}{\text{cosb}}_{2}{\text{o}}_{9}$},}\ }\href {\doibase
  10.1103/PhysRevB.91.134423} {\bibfield  {journal} {\bibinfo  {journal} {Phys.
  Rev. B}\ }\textbf {\bibinfo {volume} {91}},\ \bibinfo {pages} {134423}
  (\bibinfo {year} {2015})}\BibitemShut {NoStop}%
\bibitem [{\citenamefont {Suzuki}\ \emph {et~al.}(2014)\citenamefont {Suzuki},
  \citenamefont {Matsubara}, \citenamefont {Fujiki},\ and\ \citenamefont
  {Shirakura}}]{SuzukiPhysRevB.90.184414}%
  \BibitemOpen
  \bibfield  {author} {\bibinfo {author} {\bibfnamefont {Nobuo}\ \bibnamefont
  {Suzuki}}, \bibinfo {author} {\bibfnamefont {Fumitaka}\ \bibnamefont
  {Matsubara}}, \bibinfo {author} {\bibfnamefont {Sumiyoshi}\ \bibnamefont
  {Fujiki}}, \ and\ \bibinfo {author} {\bibfnamefont {Takayuki}\ \bibnamefont
  {Shirakura}},\ }\bibfield  {title} {\enquote {\bibinfo {title} {Absence of
  classical long-range order in an $s=\frac{1}{2}$ heisenberg antiferromagnet
  on a triangular lattice},}\ }\href {\doibase 10.1103/PhysRevB.90.184414}
  {\bibfield  {journal} {\bibinfo  {journal} {Phys. Rev. B}\ }\textbf {\bibinfo
  {volume} {90}},\ \bibinfo {pages} {184414} (\bibinfo {year}
  {2014})}\BibitemShut {NoStop}%
\bibitem [{\citenamefont {Weichselbaum}\ and\ \citenamefont
  {White}(2011)}]{WhitePhysRevB.84.245130}%
  \BibitemOpen
  \bibfield  {author} {\bibinfo {author} {\bibfnamefont {Andreas}\ \bibnamefont
  {Weichselbaum}}\ and\ \bibinfo {author} {\bibfnamefont {Steven~R.}\
  \bibnamefont {White}},\ }\bibfield  {title} {\enquote {\bibinfo {title}
  {Incommensurate correlations in the anisotropic triangular heisenberg
  lattice},}\ }\href {\doibase 10.1103/PhysRevB.84.245130} {\bibfield
  {journal} {\bibinfo  {journal} {Phys. Rev. B}\ }\textbf {\bibinfo {volume}
  {84}},\ \bibinfo {pages} {245130} (\bibinfo {year} {2011})}\BibitemShut
  {NoStop}%
\bibitem [{\citenamefont {Hauke}(2013)}]{HaukePhysRevB.87.014415}%
  \BibitemOpen
  \bibfield  {author} {\bibinfo {author} {\bibfnamefont {Philipp}\ \bibnamefont
  {Hauke}},\ }\bibfield  {title} {\enquote {\bibinfo {title} {Quantum disorder
  in the spatially completely anisotropic triangular lattice},}\ }\href
  {\doibase 10.1103/PhysRevB.87.014415} {\bibfield  {journal} {\bibinfo
  {journal} {Phys. Rev. B}\ }\textbf {\bibinfo {volume} {87}},\ \bibinfo
  {pages} {014415} (\bibinfo {year} {2013})}\BibitemShut {NoStop}%
\bibitem [{\citenamefont {Coldea}\ \emph
  {et~al.}(2001{\natexlab{b}})\citenamefont {Coldea}, \citenamefont {Hayden},
  \citenamefont {Aeppli}, \citenamefont {Perring}, \citenamefont {Frost},
  \citenamefont {Mason}, \citenamefont {Cheong},\ and\ \citenamefont
  {Fisk}}]{PhysRevLett.86.5377}%
  \BibitemOpen
  \bibfield  {author} {\bibinfo {author} {\bibfnamefont {R.}~\bibnamefont
  {Coldea}}, \bibinfo {author} {\bibfnamefont {S.~M.}\ \bibnamefont {Hayden}},
  \bibinfo {author} {\bibfnamefont {G.}~\bibnamefont {Aeppli}}, \bibinfo
  {author} {\bibfnamefont {T.~G.}\ \bibnamefont {Perring}}, \bibinfo {author}
  {\bibfnamefont {C.~D.}\ \bibnamefont {Frost}}, \bibinfo {author}
  {\bibfnamefont {T.~E.}\ \bibnamefont {Mason}}, \bibinfo {author}
  {\bibfnamefont {S.-W.}\ \bibnamefont {Cheong}}, \ and\ \bibinfo {author}
  {\bibfnamefont {Z.}~\bibnamefont {Fisk}},\ }\bibfield  {title} {\enquote
  {\bibinfo {title} {Spin waves and electronic interactions in
  ${\mathrm{la}}_{2}{\mathrm{cuo}}_{4}$},}\ }\href {\doibase
  10.1103/PhysRevLett.86.5377} {\bibfield  {journal} {\bibinfo  {journal}
  {Phys. Rev. Lett.}\ }\textbf {\bibinfo {volume} {86}},\ \bibinfo {pages}
  {5377--5380} (\bibinfo {year} {2001}{\natexlab{b}})}\BibitemShut {NoStop}%
\bibitem [{\citenamefont {R\o{}nnow}\ \emph {et~al.}(2001)\citenamefont
  {R\o{}nnow}, \citenamefont {McMorrow}, \citenamefont {Coldea}, \citenamefont
  {Harrison}, \citenamefont {Youngson}, \citenamefont {Perring}, \citenamefont
  {Aeppli}, \citenamefont {Sylju\aa{}sen}, \citenamefont {Lefmann},\ and\
  \citenamefont {Rischel}}]{PhysRevLett.87.037202}%
  \BibitemOpen
  \bibfield  {author} {\bibinfo {author} {\bibfnamefont {H.~M.}\ \bibnamefont
  {R\o{}nnow}}, \bibinfo {author} {\bibfnamefont {D.~F.}\ \bibnamefont
  {McMorrow}}, \bibinfo {author} {\bibfnamefont {R.}~\bibnamefont {Coldea}},
  \bibinfo {author} {\bibfnamefont {A.}~\bibnamefont {Harrison}}, \bibinfo
  {author} {\bibfnamefont {I.~D.}\ \bibnamefont {Youngson}}, \bibinfo {author}
  {\bibfnamefont {T.~G.}\ \bibnamefont {Perring}}, \bibinfo {author}
  {\bibfnamefont {G.}~\bibnamefont {Aeppli}}, \bibinfo {author} {\bibfnamefont
  {O.}~\bibnamefont {Sylju\aa{}sen}}, \bibinfo {author} {\bibfnamefont
  {K.}~\bibnamefont {Lefmann}}, \ and\ \bibinfo {author} {\bibfnamefont
  {C.}~\bibnamefont {Rischel}},\ }\bibfield  {title} {\enquote {\bibinfo
  {title} {Spin dynamics of the 2d spin $\frac{1}{2}$ quantum antiferromagnet
  copper deuteroformate tetradeuterate (cftd)},}\ }\href {\doibase
  10.1103/PhysRevLett.87.037202} {\bibfield  {journal} {\bibinfo  {journal}
  {Phys. Rev. Lett.}\ }\textbf {\bibinfo {volume} {87}},\ \bibinfo {pages}
  {037202} (\bibinfo {year} {2001})}\BibitemShut {NoStop}%
\bibitem [{\citenamefont {Ament}\ \emph {et~al.}(2011)\citenamefont {Ament},
  \citenamefont {van Veenendaal}, \citenamefont {Devereaux}, \citenamefont
  {Hill},\ and\ \citenamefont {van~den Brink}}]{RevModPhys.83.705}%
  \BibitemOpen
  \bibfield  {author} {\bibinfo {author} {\bibfnamefont {Luuk J.~P.}\
  \bibnamefont {Ament}}, \bibinfo {author} {\bibfnamefont {Michel}\
  \bibnamefont {van Veenendaal}}, \bibinfo {author} {\bibfnamefont {Thomas~P.}\
  \bibnamefont {Devereaux}}, \bibinfo {author} {\bibfnamefont {John~P.}\
  \bibnamefont {Hill}}, \ and\ \bibinfo {author} {\bibfnamefont {Jeroen}\
  \bibnamefont {van~den Brink}},\ }\bibfield  {title} {\enquote {\bibinfo
  {title} {Resonant inelastic x-ray scattering studies of elementary
  excitations},}\ }\href {\doibase 10.1103/RevModPhys.83.705} {\bibfield
  {journal} {\bibinfo  {journal} {Rev. Mod. Phys.}\ }\textbf {\bibinfo {volume}
  {83}},\ \bibinfo {pages} {705--767} (\bibinfo {year} {2011})}\BibitemShut
  {NoStop}%
\bibitem [{\citenamefont {Hill}\ \emph {et~al.}(2008)\citenamefont {Hill},
  \citenamefont {Blumberg}, \citenamefont {Kim}, \citenamefont {Ellis},
  \citenamefont {Wakimoto}, \citenamefont {Birgeneau}, \citenamefont {Komiya},
  \citenamefont {Ando}, \citenamefont {Liang}, \citenamefont {Greene},
  \citenamefont {Casa},\ and\ \citenamefont {Gog}}]{PhysRevLett.100.097001}%
  \BibitemOpen
  \bibfield  {author} {\bibinfo {author} {\bibfnamefont {J.~P.}\ \bibnamefont
  {Hill}}, \bibinfo {author} {\bibfnamefont {G.}~\bibnamefont {Blumberg}},
  \bibinfo {author} {\bibfnamefont {Young-June}\ \bibnamefont {Kim}}, \bibinfo
  {author} {\bibfnamefont {D.~S.}\ \bibnamefont {Ellis}}, \bibinfo {author}
  {\bibfnamefont {S.}~\bibnamefont {Wakimoto}}, \bibinfo {author}
  {\bibfnamefont {R.~J.}\ \bibnamefont {Birgeneau}}, \bibinfo {author}
  {\bibfnamefont {Seiki}\ \bibnamefont {Komiya}}, \bibinfo {author}
  {\bibfnamefont {Yoichi}\ \bibnamefont {Ando}}, \bibinfo {author}
  {\bibfnamefont {B.}~\bibnamefont {Liang}}, \bibinfo {author} {\bibfnamefont
  {R.~L.}\ \bibnamefont {Greene}}, \bibinfo {author} {\bibfnamefont
  {D.}~\bibnamefont {Casa}}, \ and\ \bibinfo {author} {\bibfnamefont
  {T.}~\bibnamefont {Gog}},\ }\bibfield  {title} {\enquote {\bibinfo {title}
  {Observation of a 500~mev collective mode in
  ${\mathrm{la}}_{2-x}{\mathrm{sr}}_{x}{\mathrm{cuo}}_{4}$ and
  ${\mathrm{nd}}_{2}{\mathrm{cuo}}_{4}$ using resonant inelastic x-ray
  scattering},}\ }\href {\doibase 10.1103/PhysRevLett.100.097001} {\bibfield
  {journal} {\bibinfo  {journal} {Phys. Rev. Lett.}\ }\textbf {\bibinfo
  {volume} {100}},\ \bibinfo {pages} {097001} (\bibinfo {year}
  {2008})}\BibitemShut {NoStop}%
\bibitem [{\citenamefont {Ellis}\ \emph {et~al.}(2010)\citenamefont {Ellis},
  \citenamefont {Kim}, \citenamefont {Hill}, \citenamefont {Wakimoto},
  \citenamefont {Birgeneau}, \citenamefont {Shvyd'ko}, \citenamefont {Casa},
  \citenamefont {Gog}, \citenamefont {Ishii}, \citenamefont {Ikeuchi},
  \citenamefont {Paramekanti},\ and\ \citenamefont {Kim}}]{PhysRevB.81.085124}%
  \BibitemOpen
  \bibfield  {author} {\bibinfo {author} {\bibfnamefont {D.~S.}\ \bibnamefont
  {Ellis}}, \bibinfo {author} {\bibfnamefont {Jungho}\ \bibnamefont {Kim}},
  \bibinfo {author} {\bibfnamefont {J.~P.}\ \bibnamefont {Hill}}, \bibinfo
  {author} {\bibfnamefont {S.}~\bibnamefont {Wakimoto}}, \bibinfo {author}
  {\bibfnamefont {R.~J.}\ \bibnamefont {Birgeneau}}, \bibinfo {author}
  {\bibfnamefont {Y.}~\bibnamefont {Shvyd'ko}}, \bibinfo {author}
  {\bibfnamefont {D.}~\bibnamefont {Casa}}, \bibinfo {author} {\bibfnamefont
  {T.}~\bibnamefont {Gog}}, \bibinfo {author} {\bibfnamefont {K.}~\bibnamefont
  {Ishii}}, \bibinfo {author} {\bibfnamefont {K.}~\bibnamefont {Ikeuchi}},
  \bibinfo {author} {\bibfnamefont {A.}~\bibnamefont {Paramekanti}}, \ and\
  \bibinfo {author} {\bibfnamefont {Young-June}\ \bibnamefont {Kim}},\
  }\bibfield  {title} {\enquote {\bibinfo {title} {Magnetic nature of the 500
  mev peak in ${\text{la}}_{2-x}{\text{sr}}_{x}{\text{cuo}}_{4}$ observed with
  resonant inelastic x-ray scattering at the $\text{Cu}\text{\,}k$-edge},}\
  }\href {\doibase 10.1103/PhysRevB.81.085124} {\bibfield  {journal} {\bibinfo
  {journal} {Phys. Rev. B}\ }\textbf {\bibinfo {volume} {81}},\ \bibinfo
  {pages} {085124} (\bibinfo {year} {2010})}\BibitemShut {NoStop}%
\bibitem [{\citenamefont {van~den Brink}(2007)}]{EPL.80.47003}%
  \BibitemOpen
  \bibfield  {author} {\bibinfo {author} {\bibfnamefont {J.}~\bibnamefont
  {van~den Brink}},\ }\bibfield  {title} {\enquote {\bibinfo {title} {The
  theory of indirect resonant inelastic x-ray scattering on magnons},}\ }\href
  {http://stacks.iop.org/0295-5075/80/i=4/a=47003} {\bibfield  {journal}
  {\bibinfo  {journal} {Europhys. Lett.}\ }\textbf {\bibinfo {volume} {80}},\
  \bibinfo {pages} {47003} (\bibinfo {year} {2007})}\BibitemShut {NoStop}%
\bibitem [{\citenamefont {Nagao}\ and\ \citenamefont
  {Igarashi}(2007)}]{PhysRevB.75.214414}%
  \BibitemOpen
  \bibfield  {author} {\bibinfo {author} {\bibfnamefont {Tatsuya}\ \bibnamefont
  {Nagao}}\ and\ \bibinfo {author} {\bibfnamefont {Jun-Ichi}\ \bibnamefont
  {Igarashi}},\ }\bibfield  {title} {\enquote {\bibinfo {title} {Two-magnon
  excitations in resonant inelastic x-ray scattering from quantum heisenberg
  antiferromagnets},}\ }\href {\doibase 10.1103/PhysRevB.75.214414} {\bibfield
  {journal} {\bibinfo  {journal} {Phys. Rev. B}\ }\textbf {\bibinfo {volume}
  {75}},\ \bibinfo {pages} {214414} (\bibinfo {year} {2007})}\BibitemShut
  {NoStop}%
\bibitem [{\citenamefont {Forte}\ \emph {et~al.}(2008)\citenamefont {Forte},
  \citenamefont {Ament},\ and\ \citenamefont {van~den
  Brink}}]{PhysRevB.77.134428}%
  \BibitemOpen
  \bibfield  {author} {\bibinfo {author} {\bibfnamefont {Filomena}\
  \bibnamefont {Forte}}, \bibinfo {author} {\bibfnamefont {Luuk J.~P.}\
  \bibnamefont {Ament}}, \ and\ \bibinfo {author} {\bibfnamefont {Jeroen}\
  \bibnamefont {van~den Brink}},\ }\bibfield  {title} {\enquote {\bibinfo
  {title} {Magnetic excitations in
  ${\mathrm{la}}_{2}\mathrm{Cu}{\mathrm{o}}_{4}$ probed by indirect resonant
  inelastic x-ray scattering},}\ }\href {\doibase 10.1103/PhysRevB.77.134428}
  {\bibfield  {journal} {\bibinfo  {journal} {Phys. Rev. B}\ }\textbf {\bibinfo
  {volume} {77}},\ \bibinfo {pages} {134428} (\bibinfo {year}
  {2008})}\BibitemShut {NoStop}%
\bibitem [{\citenamefont {Luo}\ \emph {et~al.}(2014)\citenamefont {Luo},
  \citenamefont {Datta},\ and\ \citenamefont {Yao}}]{PhysRevB.89.165103}%
  \BibitemOpen
  \bibfield  {author} {\bibinfo {author} {\bibfnamefont {Cheng}\ \bibnamefont
  {Luo}}, \bibinfo {author} {\bibfnamefont {Trinanjan}\ \bibnamefont {Datta}},
  \ and\ \bibinfo {author} {\bibfnamefont {Dao~Xin}\ \bibnamefont {Yao}},\
  }\bibfield  {title} {\enquote {\bibinfo {title} {Spectrum splitting of
  bimagnon excitations in a spatially frustrated heisenberg antiferromagnet
  revealed by resonant inelastic x-ray scattering},}\ }\href {\doibase
  10.1103/PhysRevB.89.165103} {\bibfield  {journal} {\bibinfo  {journal} {Phys.
  Rev. B}\ }\textbf {\bibinfo {volume} {89}},\ \bibinfo {pages} {165103}
  (\bibinfo {year} {2014})}\BibitemShut {NoStop}%
\bibitem [{\citenamefont {Jia}\ \emph {et~al.}(2012)\citenamefont {Jia},
  \citenamefont {Chen}, \citenamefont {Sorini}, \citenamefont {Moritz},\ and\
  \citenamefont {Devereaux}}]{NJP.14.113038}%
  \BibitemOpen
  \bibfield  {author} {\bibinfo {author} {\bibfnamefont {C.~J.}\ \bibnamefont
  {Jia}}, \bibinfo {author} {\bibfnamefont {C.-C.}\ \bibnamefont {Chen}},
  \bibinfo {author} {\bibfnamefont {A.~P.}\ \bibnamefont {Sorini}}, \bibinfo
  {author} {\bibfnamefont {B.}~\bibnamefont {Moritz}}, \ and\ \bibinfo {author}
  {\bibfnamefont {T.~P.}\ \bibnamefont {Devereaux}},\ }\bibfield  {title}
  {\enquote {\bibinfo {title} {Uncovering selective excitations using the
  resonant profile of indirect inelastic x-ray scattering in correlated
  materials: observing two-magnon scattering and relation to the dynamical
  structure factor},}\ }\href
  {http://stacks.iop.org/1367-2630/14/i=11/a=113038} {\bibfield  {journal}
  {\bibinfo  {journal} {New Journal of Physics}\ }\textbf {\bibinfo {volume}
  {14}},\ \bibinfo {pages} {113038} (\bibinfo {year} {2012})}\BibitemShut
  {NoStop}%
\bibitem [{\citenamefont {Ament}\ \emph {et~al.}(2009)\citenamefont {Ament},
  \citenamefont {Ghiringhelli}, \citenamefont {Sala}, \citenamefont
  {Braicovich},\ and\ \citenamefont {van~den Brink}}]{PhysRevLett.103.117003}%
  \BibitemOpen
  \bibfield  {author} {\bibinfo {author} {\bibfnamefont {Luuk J.~P.}\
  \bibnamefont {Ament}}, \bibinfo {author} {\bibfnamefont {Giacomo}\
  \bibnamefont {Ghiringhelli}}, \bibinfo {author} {\bibfnamefont
  {Marco~Moretti}\ \bibnamefont {Sala}}, \bibinfo {author} {\bibfnamefont
  {Lucio}\ \bibnamefont {Braicovich}}, \ and\ \bibinfo {author} {\bibfnamefont
  {Jeroen}\ \bibnamefont {van~den Brink}},\ }\bibfield  {title} {\enquote
  {\bibinfo {title} {Theoretical demonstration of how the dispersion of
  magnetic excitations in cuprate compounds can be determined using resonant
  inelastic x-ray scattering},}\ }\href {\doibase
  10.1103/PhysRevLett.103.117003} {\bibfield  {journal} {\bibinfo  {journal}
  {Phys. Rev. Lett.}\ }\textbf {\bibinfo {volume} {103}},\ \bibinfo {pages}
  {117003} (\bibinfo {year} {2009})}\BibitemShut {NoStop}%
\bibitem [{\citenamefont {Haverkort}(2010)}]{PhysRevLett.105.167404}%
  \BibitemOpen
  \bibfield  {author} {\bibinfo {author} {\bibfnamefont {M.~W.}\ \bibnamefont
  {Haverkort}},\ }\bibfield  {title} {\enquote {\bibinfo {title} {Theory of
  resonant inelastic x-ray scattering by collective magnetic excitations},}\
  }\href {\doibase 10.1103/PhysRevLett.105.167404} {\bibfield  {journal}
  {\bibinfo  {journal} {Phys. Rev. Lett.}\ }\textbf {\bibinfo {volume} {105}},\
  \bibinfo {pages} {167404} (\bibinfo {year} {2010})}\BibitemShut {NoStop}%
\bibitem [{\citenamefont {Ament}\ and\ \citenamefont {van~den
  Brink}()}]{arXiv:1002.3773}%
  \BibitemOpen
  \bibfield  {author} {\bibinfo {author} {\bibfnamefont {Luuk J.~P.}\
  \bibnamefont {Ament}}\ and\ \bibinfo {author} {\bibfnamefont {Jeroen}\
  \bibnamefont {van~den Brink}},\ }\href@noop {} {\enquote {\bibinfo {title}
  {Strong three-magnon scattering in cuprates by resonant x-rays},}\ }\Eprint
  {http://arxiv.org/abs/1002.3773} {arXiv:1002.3773} \BibitemShut {NoStop}%
\bibitem [{\citenamefont {Zhitomirsky}\ and\ \citenamefont
  {Chernyshev}(2013)}]{RevModPhys.85.219}%
  \BibitemOpen
  \bibfield  {author} {\bibinfo {author} {\bibfnamefont {M.~E.}\ \bibnamefont
  {Zhitomirsky}}\ and\ \bibinfo {author} {\bibfnamefont {A.~L.}\ \bibnamefont
  {Chernyshev}},\ }\bibfield  {title} {\enquote {\bibinfo {title} {Colloquium:
  Spontaneous magnon decays},}\ }\href {\doibase 10.1103/RevModPhys.85.219}
  {\bibfield  {journal} {\bibinfo  {journal} {Rev. Mod. Phys.}\ }\textbf
  {\bibinfo {volume} {85}},\ \bibinfo {pages} {219--242} (\bibinfo {year}
  {2013})}\BibitemShut {NoStop}%
\bibitem [{\citenamefont {Chernyshev}\ and\ \citenamefont
  {Zhitomirsky}(2006)}]{PhysRevLett.97.207202}%
  \BibitemOpen
  \bibfield  {author} {\bibinfo {author} {\bibfnamefont {A.~L.}\ \bibnamefont
  {Chernyshev}}\ and\ \bibinfo {author} {\bibfnamefont {M.~E.}\ \bibnamefont
  {Zhitomirsky}},\ }\bibfield  {title} {\enquote {\bibinfo {title} {Magnon
  decay in noncollinear quantum antiferromagnets},}\ }\href {\doibase
  10.1103/PhysRevLett.97.207202} {\bibfield  {journal} {\bibinfo  {journal}
  {Phys. Rev. Lett.}\ }\textbf {\bibinfo {volume} {97}},\ \bibinfo {pages}
  {207202} (\bibinfo {year} {2006})}\BibitemShut {NoStop}%
\bibitem [{\citenamefont {van~den Brink}\ and\ \citenamefont {van
  Veenendaal}(2006)}]{EPL.73.121}%
  \BibitemOpen
  \bibfield  {author} {\bibinfo {author} {\bibfnamefont {J.}~\bibnamefont
  {van~den Brink}}\ and\ \bibinfo {author} {\bibfnamefont {M.}~\bibnamefont
  {van Veenendaal}},\ }\bibfield  {title} {\enquote {\bibinfo {title}
  {Correlation functions measured by indirect resonant inelastic x-ray
  scattering},}\ }\href {http://stacks.iop.org/0295-5075/73/i=1/a=121}
  {\bibfield  {journal} {\bibinfo  {journal} {Europhys. Lett.}\ }\textbf
  {\bibinfo {volume} {73}},\ \bibinfo {pages} {121} (\bibinfo {year}
  {2006})}\BibitemShut {NoStop}%
\bibitem [{\citenamefont {Ament}\ \emph {et~al.}(2007)\citenamefont {Ament},
  \citenamefont {Forte},\ and\ \citenamefont {van~den
  Brink}}]{PhysRevB.75.115118}%
  \BibitemOpen
  \bibfield  {author} {\bibinfo {author} {\bibfnamefont {Luuk J.~P.}\
  \bibnamefont {Ament}}, \bibinfo {author} {\bibfnamefont {Filomena}\
  \bibnamefont {Forte}}, \ and\ \bibinfo {author} {\bibfnamefont {Jeroen}\
  \bibnamefont {van~den Brink}},\ }\bibfield  {title} {\enquote {\bibinfo
  {title} {Ultrashort lifetime expansion for indirect resonant inelastic x-ray
  scattering},}\ }\href {\doibase 10.1103/PhysRevB.75.115118} {\bibfield
  {journal} {\bibinfo  {journal} {Phys. Rev. B}\ }\textbf {\bibinfo {volume}
  {75}},\ \bibinfo {pages} {115118} (\bibinfo {year} {2007})}\BibitemShut
  {NoStop}%
\bibitem [{\citenamefont {Devereaux}\ and\ \citenamefont
  {Hackl}(2007)}]{RevModPhys.79.175}%
  \BibitemOpen
  \bibfield  {author} {\bibinfo {author} {\bibfnamefont {Thomas~P.}\
  \bibnamefont {Devereaux}}\ and\ \bibinfo {author} {\bibfnamefont {Rudi}\
  \bibnamefont {Hackl}},\ }\bibfield  {title} {\enquote {\bibinfo {title}
  {Inelastic light scattering from correlated electrons},}\ }\href {\doibase
  10.1103/RevModPhys.79.175} {\bibfield  {journal} {\bibinfo  {journal} {Rev.
  Mod. Phys.}\ }\textbf {\bibinfo {volume} {79}},\ \bibinfo {pages} {175--233}
  (\bibinfo {year} {2007})}\BibitemShut {NoStop}%
\bibitem [{\citenamefont {Vernay}\ \emph {et~al.}(2007)\citenamefont {Vernay},
  \citenamefont {Devereaux},\ and\ \citenamefont {Gingras}}]{JPCM.19.145243}%
  \BibitemOpen
  \bibfield  {author} {\bibinfo {author} {\bibfnamefont {F}~\bibnamefont
  {Vernay}}, \bibinfo {author} {\bibfnamefont {T~P}\ \bibnamefont {Devereaux}},
  \ and\ \bibinfo {author} {\bibfnamefont {M~J~P}\ \bibnamefont {Gingras}},\
  }\bibfield  {title} {\enquote {\bibinfo {title} {Raman scattering for
  triangular lattices spin-1/2 heisenberg antiferromagnets},}\ }\href
  {http://stacks.iop.org/0953-8984/19/i=14/a=145243} {\bibfield  {journal}
  {\bibinfo  {journal} {J. Phys.: Condens. Matter}\ }\textbf {\bibinfo {volume}
  {19}},\ \bibinfo {pages} {145243} (\bibinfo {year} {2007})}\BibitemShut
  {NoStop}%
\bibitem [{\citenamefont {Perkins}\ and\ \citenamefont
  {Brenig}(2008)}]{PhysRevB.77.174412}%
  \BibitemOpen
  \bibfield  {author} {\bibinfo {author} {\bibfnamefont {Natalia}\ \bibnamefont
  {Perkins}}\ and\ \bibinfo {author} {\bibfnamefont {Wolfram}\ \bibnamefont
  {Brenig}},\ }\bibfield  {title} {\enquote {\bibinfo {title} {Raman scattering
  in a heisenberg $s=\frac{1}{2}$ antiferromagnet on the triangular lattice},}\
  }\href {\doibase 10.1103/PhysRevB.77.174412} {\bibfield  {journal} {\bibinfo
  {journal} {Phys. Rev. B}\ }\textbf {\bibinfo {volume} {77}},\ \bibinfo
  {pages} {174412} (\bibinfo {year} {2008})}\BibitemShut {NoStop}%
\bibitem [{\citenamefont {Perkins}\ \emph {et~al.}(2013)\citenamefont
  {Perkins}, \citenamefont {Chern},\ and\ \citenamefont
  {Brenig}}]{PhysRevB.87.174423}%
  \BibitemOpen
  \bibfield  {author} {\bibinfo {author} {\bibfnamefont {Natalia~B.}\
  \bibnamefont {Perkins}}, \bibinfo {author} {\bibfnamefont {Gia-Wei}\
  \bibnamefont {Chern}}, \ and\ \bibinfo {author} {\bibfnamefont {Wolfram}\
  \bibnamefont {Brenig}},\ }\bibfield  {title} {\enquote {\bibinfo {title}
  {Raman scattering in a heisenberg $s=\frac{1}{2}$ antiferromagnet on the
  anisotropic triangular lattice},}\ }\href {\doibase
  10.1103/PhysRevB.87.174423} {\bibfield  {journal} {\bibinfo  {journal} {Phys.
  Rev. B}\ }\textbf {\bibinfo {volume} {87}},\ \bibinfo {pages} {174423}
  (\bibinfo {year} {2013})}\BibitemShut {NoStop}%
\bibitem [{\citenamefont {Zheng}\ \emph
  {et~al.}(2006{\natexlab{b}})\citenamefont {Zheng}, \citenamefont
  {Fj\ae{}restad}, \citenamefont {Singh}, \citenamefont {McKenzie},\ and\
  \citenamefont {Coldea}}]{PhysRevLett.96.057201}%
  \BibitemOpen
  \bibfield  {author} {\bibinfo {author} {\bibfnamefont {Weihong}\ \bibnamefont
  {Zheng}}, \bibinfo {author} {\bibfnamefont {John~O.}\ \bibnamefont
  {Fj\ae{}restad}}, \bibinfo {author} {\bibfnamefont {Rajiv R.~P.}\
  \bibnamefont {Singh}}, \bibinfo {author} {\bibfnamefont {Ross~H.}\
  \bibnamefont {McKenzie}}, \ and\ \bibinfo {author} {\bibfnamefont {Radu}\
  \bibnamefont {Coldea}},\ }\bibfield  {title} {\enquote {\bibinfo {title}
  {Anomalous excitation spectra of frustrated quantum antiferromagnets},}\
  }\href {\doibase 10.1103/PhysRevLett.96.057201} {\bibfield  {journal}
  {\bibinfo  {journal} {Phys. Rev. Lett.}\ }\textbf {\bibinfo {volume} {96}},\
  \bibinfo {pages} {057201} (\bibinfo {year} {2006}{\natexlab{b}})}\BibitemShut
  {NoStop}%
\bibitem [{\citenamefont {Starykh}\ \emph {et~al.}(2006)\citenamefont
  {Starykh}, \citenamefont {Chubukov},\ and\ \citenamefont
  {Abanov}}]{PhysRevB.74.180403}%
  \BibitemOpen
  \bibfield  {author} {\bibinfo {author} {\bibfnamefont {Oleg~A.}\ \bibnamefont
  {Starykh}}, \bibinfo {author} {\bibfnamefont {Andrey~V.}\ \bibnamefont
  {Chubukov}}, \ and\ \bibinfo {author} {\bibfnamefont {Alexander~G.}\
  \bibnamefont {Abanov}},\ }\bibfield  {title} {\enquote {\bibinfo {title}
  {Flat spin-wave dispersion in a triangular antiferromagnet},}\ }\href
  {\doibase 10.1103/PhysRevB.74.180403} {\bibfield  {journal} {\bibinfo
  {journal} {Phys. Rev. B}\ }\textbf {\bibinfo {volume} {74}},\ \bibinfo
  {pages} {180403} (\bibinfo {year} {2006})}\BibitemShut {NoStop}%
\bibitem [{\citenamefont {Chernyshev}\ and\ \citenamefont
  {Zhitomirsky}(2009)}]{PhysRevB.79.144416}%
  \BibitemOpen
  \bibfield  {author} {\bibinfo {author} {\bibfnamefont {A.~L.}\ \bibnamefont
  {Chernyshev}}\ and\ \bibinfo {author} {\bibfnamefont {M.~E.}\ \bibnamefont
  {Zhitomirsky}},\ }\bibfield  {title} {\enquote {\bibinfo {title} {Spin waves
  in a triangular lattice antiferromagnet: Decays, spectrum renormalization,
  and singularities},}\ }\href {\doibase 10.1103/PhysRevB.79.144416} {\bibfield
   {journal} {\bibinfo  {journal} {Phys. Rev. B}\ }\textbf {\bibinfo {volume}
  {79}},\ \bibinfo {pages} {144416} (\bibinfo {year} {2009})}\BibitemShut
  {NoStop}%
\bibitem [{\citenamefont {Zhou}\ \emph {et~al.}(2012)\citenamefont {Zhou},
  \citenamefont {Xu}, \citenamefont {Hallas}, \citenamefont {Silverstein},
  \citenamefont {Wiebe}, \citenamefont {Umegaki}, \citenamefont {Yan},
  \citenamefont {Murphy}, \citenamefont {Park}, \citenamefont {Qiu},
  \citenamefont {Copley}, \citenamefont {Gardner},\ and\ \citenamefont
  {Takano}}]{PhysRevLett.109.267206}%
  \BibitemOpen
  \bibfield  {author} {\bibinfo {author} {\bibfnamefont {H.~D.}\ \bibnamefont
  {Zhou}}, \bibinfo {author} {\bibfnamefont {Cenke}\ \bibnamefont {Xu}},
  \bibinfo {author} {\bibfnamefont {A.~M.}\ \bibnamefont {Hallas}}, \bibinfo
  {author} {\bibfnamefont {H.~J.}\ \bibnamefont {Silverstein}}, \bibinfo
  {author} {\bibfnamefont {C.~R.}\ \bibnamefont {Wiebe}}, \bibinfo {author}
  {\bibfnamefont {I.}~\bibnamefont {Umegaki}}, \bibinfo {author} {\bibfnamefont
  {J.~Q.}\ \bibnamefont {Yan}}, \bibinfo {author} {\bibfnamefont {T.~P.}\
  \bibnamefont {Murphy}}, \bibinfo {author} {\bibfnamefont {J.-H.}\
  \bibnamefont {Park}}, \bibinfo {author} {\bibfnamefont {Y.}~\bibnamefont
  {Qiu}}, \bibinfo {author} {\bibfnamefont {J.~R.~D.}\ \bibnamefont {Copley}},
  \bibinfo {author} {\bibfnamefont {J.~S.}\ \bibnamefont {Gardner}}, \ and\
  \bibinfo {author} {\bibfnamefont {Y.}~\bibnamefont {Takano}},\ }\bibfield
  {title} {\enquote {\bibinfo {title} {Successive phase transitions and
  extended spin-excitation continuum in the $s\mathbf{=}\frac{1}{2}$
  triangular-lattice antiferromagnet
  ${\mathrm{ba}}_{3}{\mathrm{cosb}}_{2}{\mathbf{o}}_{9}$},}\ }\href {\doibase
  10.1103/PhysRevLett.109.267206} {\bibfield  {journal} {\bibinfo  {journal}
  {Phys. Rev. Lett.}\ }\textbf {\bibinfo {volume} {109}},\ \bibinfo {pages}
  {267206} (\bibinfo {year} {2012})}\BibitemShut {NoStop}%
\bibitem [{\citenamefont {Coldea}\ \emph {et~al.}(2003)\citenamefont {Coldea},
  \citenamefont {Tennant},\ and\ \citenamefont
  {Tylczynski}}]{PhysRevB.68.134424}%
  \BibitemOpen
  \bibfield  {author} {\bibinfo {author} {\bibfnamefont {R.}~\bibnamefont
  {Coldea}}, \bibinfo {author} {\bibfnamefont {D.~A.}\ \bibnamefont {Tennant}},
  \ and\ \bibinfo {author} {\bibfnamefont {Z.}~\bibnamefont {Tylczynski}},\
  }\bibfield  {title} {\enquote {\bibinfo {title} {Extended scattering continua
  characteristic of spin fractionalization in the two-dimensional frustrated
  quantum magnet ${\mathrm{cs}}_{2}{\mathrm{cucl}}_{4}$ observed by neutron
  scattering},}\ }\href {\doibase 10.1103/PhysRevB.68.134424} {\bibfield
  {journal} {\bibinfo  {journal} {Phys. Rev. B}\ }\textbf {\bibinfo {volume}
  {68}},\ \bibinfo {pages} {134424} (\bibinfo {year} {2003})}\BibitemShut
  {NoStop}%
\bibitem [{\citenamefont {Oh}\ \emph {et~al.}(2013)\citenamefont {Oh},
  \citenamefont {Le}, \citenamefont {Jeong}, \citenamefont {Lee}, \citenamefont
  {Woo}, \citenamefont {Song}, \citenamefont {Perring}, \citenamefont {Buyers},
  \citenamefont {Cheong},\ and\ \citenamefont {Park}}]{PhysRevLett.111.257202}%
  \BibitemOpen
  \bibfield  {author} {\bibinfo {author} {\bibfnamefont {Joosung}\ \bibnamefont
  {Oh}}, \bibinfo {author} {\bibfnamefont {Manh~Duc}\ \bibnamefont {Le}},
  \bibinfo {author} {\bibfnamefont {Jaehong}\ \bibnamefont {Jeong}}, \bibinfo
  {author} {\bibfnamefont {Jung-hyun}\ \bibnamefont {Lee}}, \bibinfo {author}
  {\bibfnamefont {Hyungje}\ \bibnamefont {Woo}}, \bibinfo {author}
  {\bibfnamefont {Wan-Young}\ \bibnamefont {Song}}, \bibinfo {author}
  {\bibfnamefont {T.~G.}\ \bibnamefont {Perring}}, \bibinfo {author}
  {\bibfnamefont {W.~J.~L.}\ \bibnamefont {Buyers}}, \bibinfo {author}
  {\bibfnamefont {S.-W.}\ \bibnamefont {Cheong}}, \ and\ \bibinfo {author}
  {\bibfnamefont {Je-Geun}\ \bibnamefont {Park}},\ }\bibfield  {title}
  {\enquote {\bibinfo {title} {Magnon breakdown in a two dimensional triangular
  lattice heisenberg antiferromagnet of multiferroic ${\mathrm{lumno}}_{3}$},}\
  }\href {\doibase 10.1103/PhysRevLett.111.257202} {\bibfield  {journal}
  {\bibinfo  {journal} {Phys. Rev. Lett.}\ }\textbf {\bibinfo {volume} {111}},\
  \bibinfo {pages} {257202} (\bibinfo {year} {2013})}\BibitemShut {NoStop}%
\bibitem [{\citenamefont {Veillette}\ \emph {et~al.}(2005)\citenamefont
  {Veillette}, \citenamefont {James},\ and\ \citenamefont
  {Essler}}]{PhysRevB.72.134429}%
  \BibitemOpen
  \bibfield  {author} {\bibinfo {author} {\bibfnamefont {M.~Y.}\ \bibnamefont
  {Veillette}}, \bibinfo {author} {\bibfnamefont {A.~J.~A.}\ \bibnamefont
  {James}}, \ and\ \bibinfo {author} {\bibfnamefont {F.~H.~L.}\ \bibnamefont
  {Essler}},\ }\bibfield  {title} {\enquote {\bibinfo {title} {Spin dynamics of
  the quasi-two-dimensional spin-$\frac{1}{2}$ quantum magnet
  ${\mathrm{cs}}_{2}{\mathrm{cucl}}_{4}$},}\ }\href {\doibase
  10.1103/PhysRevB.72.134429} {\bibfield  {journal} {\bibinfo  {journal} {Phys.
  Rev. B}\ }\textbf {\bibinfo {volume} {72}},\ \bibinfo {pages} {134429}
  (\bibinfo {year} {2005})}\BibitemShut {NoStop}%
\bibitem [{\citenamefont {Dalidovich}\ \emph {et~al.}(2006)\citenamefont
  {Dalidovich}, \citenamefont {Sknepnek}, \citenamefont {Berlinsky},
  \citenamefont {Zhang},\ and\ \citenamefont {Kallin}}]{PhysRevB.73.184403}%
  \BibitemOpen
  \bibfield  {author} {\bibinfo {author} {\bibfnamefont {Denis}\ \bibnamefont
  {Dalidovich}}, \bibinfo {author} {\bibfnamefont {Rastko}\ \bibnamefont
  {Sknepnek}}, \bibinfo {author} {\bibfnamefont {A.~John}\ \bibnamefont
  {Berlinsky}}, \bibinfo {author} {\bibfnamefont {Junhua}\ \bibnamefont
  {Zhang}}, \ and\ \bibinfo {author} {\bibfnamefont {Catherine}\ \bibnamefont
  {Kallin}},\ }\bibfield  {title} {\enquote {\bibinfo {title} {Spin structure
  factor of the frustrated quantum magnet
  ${\mathrm{cs}}_{2}{\mathrm{cucl}}_{4}$},}\ }\href {\doibase
  10.1103/PhysRevB.73.184403} {\bibfield  {journal} {\bibinfo  {journal} {Phys.
  Rev. B}\ }\textbf {\bibinfo {volume} {73}},\ \bibinfo {pages} {184403}
  (\bibinfo {year} {2006})}\BibitemShut {NoStop}%
\bibitem [{\citenamefont {Mourigal}\ \emph {et~al.}(2013)\citenamefont
  {Mourigal}, \citenamefont {Fuhrman}, \citenamefont {Chernyshev},\ and\
  \citenamefont {Zhitomirsky}}]{PhysRevB.88.094407}%
  \BibitemOpen
  \bibfield  {author} {\bibinfo {author} {\bibfnamefont {M.}~\bibnamefont
  {Mourigal}}, \bibinfo {author} {\bibfnamefont {W.~T.}\ \bibnamefont
  {Fuhrman}}, \bibinfo {author} {\bibfnamefont {A.~L.}\ \bibnamefont
  {Chernyshev}}, \ and\ \bibinfo {author} {\bibfnamefont {M.~E.}\ \bibnamefont
  {Zhitomirsky}},\ }\bibfield  {title} {\enquote {\bibinfo {title} {Dynamical
  structure factor of the triangular-lattice antiferromagnet},}\ }\href
  {\doibase 10.1103/PhysRevB.88.094407} {\bibfield  {journal} {\bibinfo
  {journal} {Phys. Rev. B}\ }\textbf {\bibinfo {volume} {88}},\ \bibinfo
  {pages} {094407} (\bibinfo {year} {2013})}\BibitemShut {NoStop}%
\bibitem [{\citenamefont {Canali}\ and\ \citenamefont
  {Wallin}(1993)}]{PhysRevB.48.3264}%
  \BibitemOpen
  \bibfield  {author} {\bibinfo {author} {\bibfnamefont {C.~M.}\ \bibnamefont
  {Canali}}\ and\ \bibinfo {author} {\bibfnamefont {Mats}\ \bibnamefont
  {Wallin}},\ }\bibfield  {title} {\enquote {\bibinfo {title} {Spin-spin
  correlation functions for the square-lattice heisenberg antiferromagnet at
  zero temperature},}\ }\href {\doibase 10.1103/PhysRevB.48.3264} {\bibfield
  {journal} {\bibinfo  {journal} {Phys. Rev. B}\ }\textbf {\bibinfo {volume}
  {48}},\ \bibinfo {pages} {3264--3280} (\bibinfo {year} {1993})}\BibitemShut
  {NoStop}%
\bibitem [{\citenamefont {Lorenzana}\ \emph {et~al.}(2005)\citenamefont
  {Lorenzana}, \citenamefont {Seibold},\ and\ \citenamefont
  {Coldea}}]{PhysRevB.72.224511}%
  \BibitemOpen
  \bibfield  {author} {\bibinfo {author} {\bibfnamefont {J.}~\bibnamefont
  {Lorenzana}}, \bibinfo {author} {\bibfnamefont {G.}~\bibnamefont {Seibold}},
  \ and\ \bibinfo {author} {\bibfnamefont {R.}~\bibnamefont {Coldea}},\
  }\bibfield  {title} {\enquote {\bibinfo {title} {Sum rules and missing
  spectral weight in magnetic neutron scattering in the cuprates},}\ }\href
  {\doibase 10.1103/PhysRevB.72.224511} {\bibfield  {journal} {\bibinfo
  {journal} {Phys. Rev. B}\ }\textbf {\bibinfo {volume} {72}},\ \bibinfo
  {pages} {224511} (\bibinfo {year} {2005})}\BibitemShut {NoStop}%
\bibitem [{\citenamefont {Davies}\ \emph {et~al.}(1971)\citenamefont {Davies},
  \citenamefont {Chinn},\ and\ \citenamefont {Zeiger}}]{PhysRevB.4.992}%
  \BibitemOpen
  \bibfield  {author} {\bibinfo {author} {\bibfnamefont {R.~W.}\ \bibnamefont
  {Davies}}, \bibinfo {author} {\bibfnamefont {S.~R.}\ \bibnamefont {Chinn}}, \
  and\ \bibinfo {author} {\bibfnamefont {H.~J.}\ \bibnamefont {Zeiger}},\
  }\bibfield  {title} {\enquote {\bibinfo {title} {Spin-wave approach to
  two-magnon raman scattering in a simple antiferromagnet},}\ }\href {\doibase
  10.1103/PhysRevB.4.992} {\bibfield  {journal} {\bibinfo  {journal} {Phys.
  Rev. B}\ }\textbf {\bibinfo {volume} {4}},\ \bibinfo {pages} {992--1004}
  (\bibinfo {year} {1971})}\BibitemShut {NoStop}%
\bibitem [{\citenamefont {Canali}\ and\ \citenamefont
  {Girvin}(1992)}]{PhysRevB.45.7127}%
  \BibitemOpen
  \bibfield  {author} {\bibinfo {author} {\bibfnamefont {C.~M.}\ \bibnamefont
  {Canali}}\ and\ \bibinfo {author} {\bibfnamefont {S.~M.}\ \bibnamefont
  {Girvin}},\ }\bibfield  {title} {\enquote {\bibinfo {title} {Theory of raman
  scattering in layered cuprate materials},}\ }\href {\doibase
  10.1103/PhysRevB.45.7127} {\bibfield  {journal} {\bibinfo  {journal} {Phys.
  Rev. B}\ }\textbf {\bibinfo {volume} {45}},\ \bibinfo {pages} {7127--7160}
  (\bibinfo {year} {1992})}\BibitemShut {NoStop}%
\bibitem [{\citenamefont {Ko}\ and\ \citenamefont
  {Lee}(2011)}]{winghoPhysRevB.84.125102}%
  \BibitemOpen
  \bibfield  {author} {\bibinfo {author} {\bibfnamefont {Wing-Ho}\ \bibnamefont
  {Ko}}\ and\ \bibinfo {author} {\bibfnamefont {Patrick~A.}\ \bibnamefont
  {Lee}},\ }\bibfield  {title} {\enquote {\bibinfo {title} {Proposal for
  detecting spin-chirality terms in mott insulators via resonant inelastic
  x-ray scattering},}\ }\href {\doibase 10.1103/PhysRevB.84.125102} {\bibfield
  {journal} {\bibinfo  {journal} {Phys. Rev. B}\ }\textbf {\bibinfo {volume}
  {84}},\ \bibinfo {pages} {125102} (\bibinfo {year} {2011})}\BibitemShut
  {NoStop}%
\end{thebibliography}%

\end{document}